# Qualify-Then-Borrow: A Five-Step Framework for Bayesian Borrowing Beyond Outcome Agreement

**Haitao Pan, PhD**

Department of Biostatistics, St. Jude Children's Research Hospital, Memphis, TN, USA

Email: Haitao.Pan@stjude.org

**Abstract**

**PURPOSE**

Dynamic Bayesian borrowing methods can adapt the contribution of external information according to its agreement or disagreement with current data, while existing external-control frameworks emphasize scientific assessment of the source before borrowing. What is less explicit is how those assessments should determine what external information reaches the borrowing model and when observed outcome agreement should be used to determine borrowing strength. Qualify-Then-Borrow (QTB) organizes these decisions into a prespecified framework in which identified material differences are addressed before dynamic borrowing is applied to the information that remains.

**METHODS**

QTB organizes Bayesian borrowing into a five-step decision sequence: define the target; assess the external data; classify them as qualified, repairable, or not qualified; use them directly, repair them before use, or exclude them; and then apply the selected Bayesian borrowing method. Dynamic borrowing is therefore the final step of QTB and operates on residual empirical compatibility—the agreement or disagreement that remains after identified material differences have been addressed. We evaluated QTB in 82 prespecified scenarios with 10,000 replications each. Four analyses were compared: trial-only inference, full pooling, robust-mixture borrowing without QTB, and QTB using the same robust mixture prior wherever borrowing was allowed.

**RESULTS**

Observed outcome data alone could not determine the appropriate borrowing decision. Two settings with the same external outcome distribution—one with no identified problem requiring repair or exclusion and one with known, nonrepairable endpoint misclassification—produced essentially the same robust-mixture borrowing behavior but different QTB decisions. When the misclassified external data were borrowed, full pooling produced severe distortion: 95% coverage fell to 60.8% despite a Type I error of only 0.02%. Robust-mixture downweighting substantially reduced the distortion but did not eliminate it: 95% coverage remained at 91.8%, with a −4.8 percentage-point bias. For a measured population shift, repair before borrowing reduced absolute bias from 1.94 to 0.80 percentage points at the largest external sample size. Under strong negative residual drift, the RMP reduced the posterior weight on the historical component, yet Type I error reached 10.0% in a design calibrated to a one-sided 2.5% target. When the external and current data were fully compatible, QTB reduced to the usual robust-mixture borrowing analysis and preserved its efficiency gains.

**CONCLUSION**

QTB does not introduce a new Bayesian prior or compatibility statistic. It provides a five-step decision architecture that determines whether external information proceeds unchanged, requires repair, or is excluded before dynamic borrowing begins. Dynamic borrowing then operates on the residual empirical compatibility of the information that remains. In this way, QTB specifies both the sequence of decisions that should precede borrowing and the residual empirical compatibility on which borrowing should operate. Qualification does not certify exchangeability, and dynamic downweighting remains a safeguard rather than a guarantee against residual bias.

## Introduction

Modern Bayesian borrowing methods no longer require a simple choice between fully pooling external information and ignoring it altogether. Robust mixture priors, robust MAP approaches, commensurate priors, power priors, and related methods allow the contribution of external information to adapt to empirical compatibility, commensurability, or assumptions about source bias. [1,2,10] These methods differ in how that adaptation is achieved, but they share a common role: once external information is being considered for borrowing, they determine how strongly it should contribute to the current analysis. A separate question arises earlier: what should happen to identified scientific differences before that borrowing mechanism is applied?

That earlier decision cannot be resolved by outcome agreement alone, because some of the information needed to make it is not contained in the observed outcomes. Two external sources can show similar outcome values while differing in ways that matter for the target analysis. For example, pediatric oncology studies may report similar progression-free or event-free survival while differing in event definition, time origin, censoring rules, assessment schedule, or handling of intercurrent events. In cellular therapy, one study may define follow-up from enrollment or leukapheresis while another begins at infusion, thereby excluding patients who progress during bridging therapy, experience manufacturing failure, or never receive an infusion. Likewise, similar objective response rates can arise under different requirements for measurable disease, response confirmation, or tumor-assessment schedules. These differences concern what evidence the external source represents for the target question, not simply how closely its observed outcomes agree with the current data. They therefore need to be addressed before outcome agreement is used to determine how much information to borrow.

Not all differences between external and current data represent the same problem, and they should not all be handled by simply borrowing less. Three situations should be distinguished. First, some known differences cannot be adequately repaired with the available data. For example, a CAR-T trial may define event-free survival from enrollment or leukapheresis and include patients who progress during bridging therapy, experience manufacturing failure, or never receive infusion. An external registry may instead start follow-up at infusion, so its time zero does not match that of the trial. If the earlier treatment course cannot be recovered from the source data, the target estimand cannot be reconstructed, and the source should not enter primary borrowing. Second, some known differences are repairable. An external control may contain a higher proportion of high-risk patients than the target trial, but if the relevant prognostic factors are measured in both sources and there is adequate overlap, adjustment or standardization to the target population may be possible before borrowing. Third, no identified material difference may require repair or exclusion, yet the external and current outcomes may still disagree. Reaching this stage does not establish that the two sources are exchangeable. Unrecognized differences may remain because of temporal drift, site-to-site variation, unmeasured prognostic factors, or other source effects. This is the point at which dynamic borrowing enters the QTB sequence: Step 5 uses the residual empirical compatibility that remains after known material differences have been addressed to determine how much external information should contribute. These are different problems requiring different actions—not simply different degrees of prior-data conflict.

The idea that external data should be assessed before borrowing is not new. Existing frameworks consider alignment of the target estimand, eligibility and time zero, treatment and follow-up, endpoint definition and ascertainment, measured prognostic factors and overlap, data quality, and potential sources of bias before external information is incorporated. [3–7,11,12] QTB builds on this work by organizing these assessments into a prespecified decision sequence with explicit consequences for borrowing. Based on the identified material differences, external information proceeds unchanged, undergoes a prespecified repair before use, or is excluded from primary borrowing. These decisions determine the form of external information that reaches Step 5; they do not determine the strength of borrowing itself. Dynamic borrowing then operates as the final step of QTB, using the residual empirical compatibility of the information that remains to determine how much it should contribute. Qualification therefore does not certify exchangeability; it defines what information is allowed to reach the borrowing stage and in what form.

Once Steps 1–4 have determined what external information proceeds to borrowing—either unchanged or after a prespecified repair—the external and current data may still agree or disagree. QTB calls this remaining agreement or disagreement residual empirical compatibility. "Residual" means that the comparison is made only after identified material differences have been addressed. Step 5 then uses this remaining compatibility to determine the strength of borrowing. QTB does not require a new compatibility statistic or a particular Bayesian model for this step. Outcome-adaptive methods may use the remaining agreement or disagreement directly, whereas more recent bias-aware approaches can represent residual source differences through explicit bias parameters or bias functions. [9,13] The role of QTB is to define what information reaches this final borrowing step and in what form; the selected Bayesian method then determines how residual incompatibility affects borrowing.

QTB puts this logic into five steps: define the target; assess the external data; classify them as qualified, repairable, or not qualified; use them directly, repair them before use, or exclude them; and then apply the selected Bayesian borrowing method. The individual elements of these steps are not themselves new; target definition, source assessment, alignment, adjustment, and bias evaluation are all well established. The contribution of QTB is to organize these elements into a prespecified decision architecture with explicit consequences for borrowing. Steps 1–4 determine what external information, if any, reaches the borrowing stage and in what form; Step 5 then uses the residual empirical compatibility of the information that remains to determine how much external information should contribute.

We evaluated this framework in a deliberately simple simulation setting designed to isolate its key decisions. The study asked whether observed outcome data alone can determine an appropriate borrowing decision; what happens when external data with a known, nonrepairable endpoint problem are nevertheless borrowed; whether a measured population difference should be repaired before borrowing; whether residual bias can remain after qualification when dynamic borrowing responds to disagreement; and whether QTB preserves the efficiency gains of borrowing when the external and current data are fully compatible.

## Methods

### The five-step Qualify-Then-Borrow framework

QTB is not a new Bayesian prior or borrowing model. It is a five-step decision framework that organizes how external information is evaluated and handled before the dynamic borrowing step and defines when borrowing should begin within that sequence. Steps 1–4 define the target, assess the external data, and classify and handle them as qualified, repairable, or not qualified for primary borrowing. These decisions are made without using observed outcome agreement between the external and current data to determine qualification. Step 5 is the dynamic borrowing step: the selected Bayesian borrowing method operates on the residual empirical compatibility that remains after identified material differences have been addressed and uses that remaining agreement or disagreement to determine how much external information should contribute. Figure 1 summarizes the five steps, and Table 1 summarizes the three QTB classifications and the corresponding action under each.

**Step 1: Define the target.** This step follows established estimand and target-trial principles [8,14,15]. Specify the target estimand before assessing the external source, including the target population, treatment strategy or condition of interest, outcome, handling of relevant intercurrent events, and the population-level summary measure or treatment contrast, as applicable. The target should be defined on scientific grounds before the external source is evaluated for borrowing and should not be altered simply to improve apparent compatibility. For time-to-event outcomes, specify the time origin, event definition, relevant time horizon, and censoring conventions; for repeatedly assessed outcomes, specify the assessment schedule or window, measurement requirements, and any response-confirmation rules relevant to the endpoint. All subsequent assessment of the external data is made relative to this prespecified target.

**Step 2: Assess the external data.** Relative to the prespecified target, assess the external source using information that does not depend on how similar the observed external and current outcomes happen to be. The assessment should examine alignment of the estimand and control strategy; eligibility, time zero, and follow-up; outcome definition, timing, ascertainment, and measurement; comparability with and support for the target population; availability of the prognostic factors needed for adjustment; data quality and provenance; and other design or data limitations that could affect the target analysis. For each identified material difference, document whether the information needed to address it is available and what prespecified adjustment, reconstruction, or other repair, if any, would be possible. Observed outcome agreement is not used to resolve these pre-borrowing assessments.

**Step 3: Classify the external source.** Based on the Step 2 assessment, classify the external source according to what must happen before it can enter primary borrowing. A source is qualified when no identified material difference requires a change in how the external information is constructed or analyzed before borrowing. A source is repairable when one or more material differences require such a change, but each can be addressed using the available data and a prespecified, scientifically justified repair. A source is not qualified for primary borrowing when at least one material difference cannot be adequately addressed with the available data.

QTB does not provide a universal numerical threshold for determining whether a difference is adequately repairable. That determination depends on the target estimand, the nature of the difference, the information available in the external source, and the assumptions required by the proposed repair. The basis for classification should therefore be specified and documented before observed outcome agreement is used to determine borrowing, including the identified material difference, the proposed repair, the data required to support it, and any assumptions or diagnostics needed to justify its use. Classification determines what action is taken before Step 5; it does not establish that the external and current data are exchangeable.

For example, external data with no identified material difference requiring modification may be qualified for direct borrowing. A measured difference in patient mix may be repairable when the relevant prognostic factors are available and there is sufficient overlap to support a prespecified adjustment or standardization. In contrast, external data that cannot reconstruct the target time zero or omit patients required by the target estimand would be not qualified for primary borrowing if the missing information cannot be recovered. Observed outcome agreement is not used to resolve these classifications.

**Step 4: Use directly, repair before use, or exclude.** Step 4 translates the Step 3 classification into the external information and analysis route that are passed to Step 5. Qualified external data proceed without a QTB-specific repair. Repairable external data proceed only through the specified repair. Data not qualified for primary borrowing contribute no external information to Step 5; in the present demonstration, the primary analysis then uses the randomized-trial data alone.

Repair does not necessarily require creating a new "corrected" external dataset. It may instead change how the external information is reconstructed, standardized, or represented for the target analysis. For example, if the external source contains a different proportion of high-risk patients than the target population but the relevant prognostic factor is measured with adequate overlap, the repair can preserve the external information within prognostic strata and require Step 5 to compare external and current controls within the same strata. In the population-shift scenario examined later in the simulation study, the target population is 50% in each of two prognostic strata; the repair therefore carries forward the stratum-specific external information together with the target-population weights. Step 5 then applies the same borrowing model within each stratum, after which the stratum-specific posterior quantities are combined according to the target population distribution.

Other repairs may take different forms. If an external registry ordinarily starts follow-up at treatment but contains the earlier dates and outcome information needed to reconstruct the target trial time zero, follow-up can be reconstructed accordingly before Step 5. If endpoint definitions or assessment schedules differ but the underlying component-level measurements or assessment dates are available, the endpoint may be reconstructed according to the target definition or assessment window. Similarly, measured differences in patient mix may be addressed through weighting, standardization, or outcome modeling when the required covariates and adequate overlap are available. These examples are illustrative rather than exhaustive; the repair should be specified and scientifically justified before Step 5 and supported by the information available in the external source, rather than chosen according to observed outcome agreement.

**Clinically grounded examples of repairability in oncology.**

A neuroblastoma endpoint example makes the distinction concrete. Consider a hypothetical relapsed/refractory neuroblastoma trial whose target endpoint is Cycle-2 objective response rate (ORR), while a historical NANT-like source reports best response through Cycle 4. If the historical source retains the Cycle-2 assessment dates and component-level disease information needed to apply the locked target response rule, the historical endpoint can be re-adjudicated under the Cycle-2 definition before borrowing. If only the final best-response-through-Cycle-4 indicator remains, the same apparent mismatch may not be adequately repairable. After reconstruction, any remaining difference between historical and current Cycle-2 ORR is residual empirical compatibility for Step 5.

The same principle applies to time zero. If a cellular-therapy registry is usually summarized from infusion but retains enrollment or leukapheresis dates, bridging-period events, manufacturing failures, pre-infusion progression, and never-infused patients, the target risk set and time zero may be reconstructed. If the source contains only infused patients and the omitted pre-infusion course cannot be recovered, the target estimand cannot be reconstructed and the source may be not qualified for primary borrowing for that target.

Population composition provides a third example. In a relapsed/refractory neuroblastoma setting, a historical source may contain a larger proportion of primary refractory patients than the current trial. If refractory/relapsed status is measured in both sources and there is adequate support in both strata, the repair can require borrowing within refractory and relapsed strata followed by standardization to the current trial population. This is the same logic used in World 2, but expressed in a clinically recognizable setting.

These examples show that repairability is not a fixed property of a mismatch label. It depends jointly on the target, the external source, and the information actually retained for repair. Supplementary Appendix A provides worked pediatric-oncology examples with illustrative numbers, and Supplementary Appendix B provides a practical template for documenting the QTB decision path in real applications.

Let $D_E^*$ denote the external information passed from Step 4 to Step 5:

$$D_E^* = \begin{cases} D_E, & \text{if the source is qualified,} \\ R(D_E;T), & \text{if the source is repairable,} \\ \emptyset, & \text{if the source is not qualified for primary borrowing.} \end{cases}$$

Here, $D_E$ denotes the original external data, $T$ denotes the prespecified target, and $R(D_E;T)$ denotes the prespecified repair or target-aligned representation through which the external information is allowed to enter Step 5. The symbol $\emptyset$ indicates that no external information is passed to primary borrowing. Step 5 then applies the selected Bayesian borrowing method to $D_E^*$, so that borrowing operates on the external information that remains after any required repair rather than on the original external data.

**Step 5: Determine how much information to borrow.** Once Steps 1–4 have determined what external information, if any, proceeds to borrowing, Step 5 applies the selected Bayesian borrowing method to that information. Let $D_E^*$ denote the external information passed forward from Step 4 and $D_C$ the current-control data. Step 5 determines how strongly $D_E^*$ contributes after identified material differences have been addressed. QTB calls the remaining agreement or disagreement between the external and current information **residual empirical compatibility**.

For qualified external data,

$$D_E^* = D_E,$$

so Step 5 operates on the external data as observed. When repair is required,

$$D_E^* = R(D_E;T),$$

so Step 5 operates on the repaired or target-aligned representation defined in Step 4 rather than on the original external data. Residual empirical compatibility is not a new compatibility statistic. It defines the stage at which empirical compatibility becomes relevant and the external information on which the selected borrowing method operates.

QTB does not introduce a new Bayesian borrowing model. Different Bayesian methods may translate residual compatibility or residual source uncertainty into borrowing in different ways. Outcome-adaptive methods may use the remaining outcome agreement or disagreement directly, whereas bias-aware approaches may represent residual source differences through explicit bias parameters or bias functions. In this study, Step 5 is implemented using a dynamic robust mixture prior, which uses the remaining outcome agreement to determine how much external information contributes; its specific updating rule is described below.

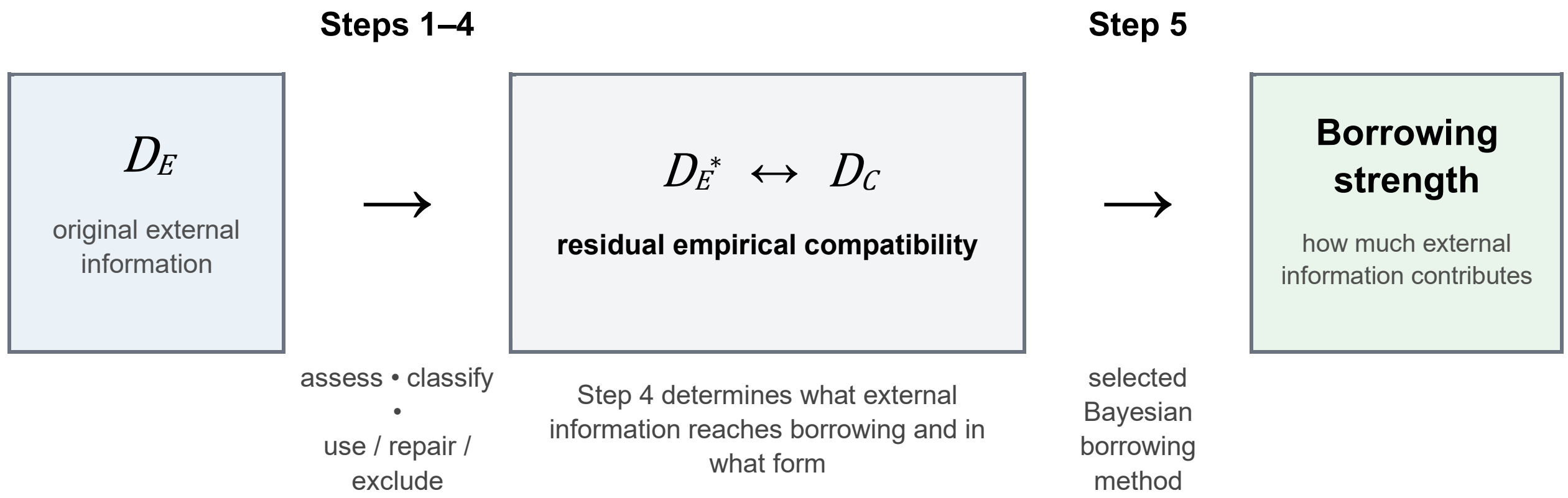

**Demonstration setting and Bayesian analyses**

We used a deliberately simple setting so that the consequences of qualification, repair, exclusion, and subsequent borrowing could be seen separately. The goal was to illustrate and stress-test the QTB decision framework rather than to reproduce a particular clinical trial. The demonstration included one randomized trial, one external-control dataset, one binary outcome, and one binary prognostic baseline covariate.

The randomized trial included 50 participants assigned to the experimental treatment and 25 concurrent controls. The randomized-trial population defined the target population, with binary baseline covariate

$$X \sim \text{Bernoulli}(0.5).$$

Let $p_T$ and $p_C$ denote the marginal treatment and control response probabilities in this target population. The target estimand was the trial-population marginal risk difference

$$\Delta_R = p_T - p_C.$$

**By design, external data were allowed to inform only the control response probability $p_C$.** They could not change the target population, outcome, treatment contrast, or estimand. Thus, all external-data scenarios were evaluated relative to the same target defined by the randomized trial.

The binary baseline covariate was chosen to create a simple prognostic structure and, in selected scenarios, a measured population difference that could be repaired before borrowing. Among current controls, $X = 1$ doubled the odds of response relative to $X = 0$, corresponding to

$$\beta_X = \log(2)$$

in the logistic model. The intercept was chosen so that the overall control response probability was $p_C = 0.30$. Under the alternative, the treatment coefficient was chosen to give an overall treatment response probability of $p_T = 0.50$, yielding

$$\Delta_R = 0.20.$$

Under the null, the treatment coefficient was zero, so

$$p_T = p_C = 0.30$$

and

$$\Delta_R = 0.$$

**Four analyses were prespecified so that the effect of QTB could be evaluated while holding the Bayesian borrowing model fixed.**

**M0** used only the randomized-trial data, with independent $\text{Beta}(1,1)$priors for the treatment and control response probabilities. **M1** fully pooled the external and concurrent control data. **M2** applied the prespecified robust mixture prior (RMP) directly to the original external controls, without the qualification and repair steps of QTB. For $y_E$responses among $n_E$external controls,

$$\pi(p_C \mid D_E) = w_0 \text{Beta}(1, +y_E \ \ 1 + n_E - y_E) + (1 - w_0)\text{Beta}(1, 1),$$

with

$$w_0 = 0.50.$$

The same fixed prior mixture weight $w_0$was used in all RMP analyses. The posterior historical-component weight could nevertheless change after updating with the concurrent-control data.

**M3** used the same RMP as M2, but applied it to the external information passed forward by QTB in Step 4, $D_E^*$. If the external data were qualified, $D_E^* = D_E$, and M3 was identical to M2. If repair was required, the same RMP was applied through the specified repair analysis. If the external source was not qualified for primary borrowing, $D_E^* = \emptyset$, no external information entered Step 5, and M3 reduced to M0. External data informed only the control response probability; the treatment arm always used randomized-trial data alone. Thus, any difference between M2 and M3 reflects the QTB decision about what external information reaches Step 5 and in what form, not a difference in the Bayesian borrowing model itself.

## M2–M3 comparison diagram

| M2 | M3 |
|---|---|
| $D_E \rightarrow$ RMP<br>Apply the RMP directly to the original external data | $D_E \rightarrow$ QTB Steps 1–4 $\rightarrow D_E^* \rightarrow$ same RMP<br>Apply the same RMP to the external information after qualification / repair |

**Key comparison:** any difference between M2 and M3 reflects what external information reaches Step 5 and in what form, not a different Bayesian borrowing model.

**Compact inline version:** M2: D_E → RMP  versus  M3: D_E → QTB Steps 1–4 → D_E* → same RMP

**How the RMP implements Step 5 of QTB**

Step 5 uses the selected Bayesian borrowing method to determine how much the external information passed forward from Steps 1–4 should contribute. In the present demonstration, that method is the robust mixture prior (RMP). QTB does not change the RMP updating rule. It changes the external information to which that rule is applied: the RMP operates on the information that remains after qualification and any required repair.

For a qualified source, QTB does not modify the external data. Let $y_E$denote the number of responses among $n_E$external controls. The historical component of the RMP is

$$p_C \mid D_E, \text{historical component} \sim \text{Beta}(a_H, b_H),$$

where

$$a_H = 1 + y_E,\ b_H = 1 + n_E - y_E.$$

The second component is the vague prior

$$p_C \sim \text{Beta}(1, 1).$$

The prespecified prior mixture is therefore

$$\pi(p_C \mid D_E) = w_0 \text{Beta}(a_H, b_H) + (1 - w_0)\text{Beta}(1, 1),$$

with

$$w_0 = 0.50.$$

After observing $y_C$responses among $n_C$concurrent controls, the current-control data update both the historical and vague components. Let $L_H$denote the posterior predictive probability of the current-control data under the historical component and $L_V$the corresponding probability under the vague component:

$$L_H = \binom{n_C}{y_C} \frac{B(a_H, +y_C\ b_H + n_C - y_C)}{B(a_H, b_H)},$$

and

$$L_V = \binom{n_C}{y_C} \frac{B(1, +y_C\ 1 + n_C - y_C)}{B(1, 1)},$$

where $B(\cdot,\cdot)$denotes the beta function. The posterior weight assigned to the historical component is

$$w_{\text{post}} = \frac{w_0 L_H}{w_0 L_H + (1 - w_0) L_V}.$$

With $w_0 = 0.50$, if the historical and vague components predict the concurrent-control data equally well, the posterior historical-component weight remains 0.50. If the historical component predicts the current data better than the vague component, its posterior weight increases; if it predicts the current data less well, its weight decreases. Thus, the RMP determines how strongly the historical information contributes through the relative predictive support provided by the current data.

The resulting posterior distribution for the control response probability is

$$\begin{aligned} p_C \mid D_E, D_C \sim\ & w_{\text{post}}\, \text{Beta}(a_H, +y_C\ b_H + n_C - y_C) \\ & + (1 - w_{\text{post}})\, \text{Beta}(1, +y_C\ 1 + n_C - y_C). \end{aligned}$$

The first posterior component combines the external and concurrent control information, whereas the second uses the concurrent controls alone. The posterior mixture weight therefore describes how strongly the historical component contributes to the posterior, but it should not be interpreted as the literal proportion of external participants that have been borrowed.

For a qualified source, this is the usual RMP update. QTB adds no new compatibility statistic or compatibility test at this stage. What QTB changes is the external information to which the RMP is applied. In this RMP implementation, the relative predictive support provided by the current data operationalizes residual empirical compatibility after identified material differences have been addressed in Steps 1–4.

For a repairable source, the same RMP is applied through the specified repair analysis. Consider the population-shift scenario examined later in the simulation study. Let $x \in \{0, 1\}$index the two levels of the prognostic baseline covariate. Within stratum $x$, let $n_{E,x}$and $y_{E,x}$denote the external sample size and number of responses, and let $n_{C,x}$and $y_{C,x}$denote the corresponding concurrent-control values. Define

$$a_{H,x} = 1 + y_{E,x},\ b_{H,x} = 1 + n_{E,x} - y_{E,x}.$$

The historical and vague predictive probabilities for the concurrent-control data in stratum $x$are

$$L_{H,x} = \binom{n_{C,x}}{y_{C,x}} \frac{B\left(a_{H,x}, +y_{C,x}\ b_{H,x} + n_{C,x} - y_{C,x}\right)}{B\left(a_{H,x}, b_{H,x}\right)},$$

and

$$L_{V,x} = \binom{n_{C,x}}{y_{C,x}} \frac{B\left(1, +y_{C,x}\ 1 + n_{C,x} - y_{C,x}\right)}{B(1, 1)}.$$

The posterior historical-component weight in stratum $x$is therefore

$$w_{\text{post},x} = \frac{w_0 L_{H,x}}{w_0 L_{H,x} + (1 - w_0) L_{V,x}}.$$

The posterior control response probability within stratum $x$is

$$\begin{aligned} p_{C,x} \mid D \sim\ & w_{\text{post},x} \operatorname{Beta}\left(1, +y_{E,x} + y_{C,x}\ 1 + n_{E,x} - y_{E,x} + n_{C,x} - y_{C,x}\right) \\ & +\left(1 - w_{\text{post},x}\right) \operatorname{Beta}\left(1, +y_{C,x}\ 1 + n_{C,x} - y_{C,x}\right). \end{aligned}$$

Thus, $X = 0$and $X = 1$each have their own RMP update and their own posterior control response probability. External and current controls are compared only within the same covariate stratum. This is the repair in the population-shift setting: the known difference in population composition determines the comparison and standardization structure before residual empirical compatibility is used to determine borrowing strength.

On posterior draw $b$, let $p_{C0}^{(b)}$and $p_{C1}^{(b)}$denote draws from the two stratum-specific control posteriors. Because the target trial population is 50% $X = 0$and 50% $X = 1$, the target-population control response draw is

$$p_C^{R,(b)} = 0.5 p_{C0}^{(b)} + 0.5 p_{C1}^{(b)}.$$

The treatment arm always uses randomized-trial data alone. If $y_T$responses are observed among $n_T$treated participants, then

$$p_T \mid D_T \sim \operatorname{Beta}(1, +y_T\ 1 + n_T - y_T).$$

Let $p_T^{(b)}$denote a draw from this posterior. The corresponding target-population treatment-effect draw after repair is

$$\Delta_R^{(b)} = p_T^{(b)} - p_C^{R,(b)}.$$

If a covariate stratum contains no external controls, that stratum receives no external borrowing and uses the concurrent-control-only posterior. If neither the external nor concurrent-control data contain observations in a stratum, the posterior for that stratum remains $\text{Beta}(1, 1)$. No external information is transferred between covariate strata.

For a source that is not qualified for primary borrowing, no external information is passed to Step 5. The control analysis therefore reduces to the concurrent-control-only posterior,

$$p_C \mid D_C \sim \text{Beta}(1, +y_C \ \ 1 + n_C - y_C),$$

and the resulting QTB analysis is identical to the randomized-trial-only analysis M0.

The schematic below summarizes the distinction between marginal borrowing and QTB borrowing in the repairable population-shift setting. Without repair, the borrowing model compares the marginal external and concurrent-control outcomes directly. Under QTB, the known population difference is addressed first: external and concurrent controls are compared within the same prognostic strata, the same RMP determines borrowing within each stratum, and the resulting posterior quantities are then standardized to the target population. In short, QTB first aligns the information being compared and then allows the selected Bayesian method to determine how much of that information to borrow.

Supplementary Algorithm S1 provides a step-by-step implementation of the binary-endpoint QTB workflow, including minimal R code for the qualified, repairable, and excluded routes. Supplementary Illustrations 1 and 2 show how the same routing logic extends to a simple continuous endpoint using a Normal-Normal robust mixture analysis and to a lightweight time-to-event endpoint using an exponential-Gamma robust mixture analysis. These illustrations are implementation examples only and were not part of the simulation study.

**How QTB repair changes the information used for dynamic borrowing**

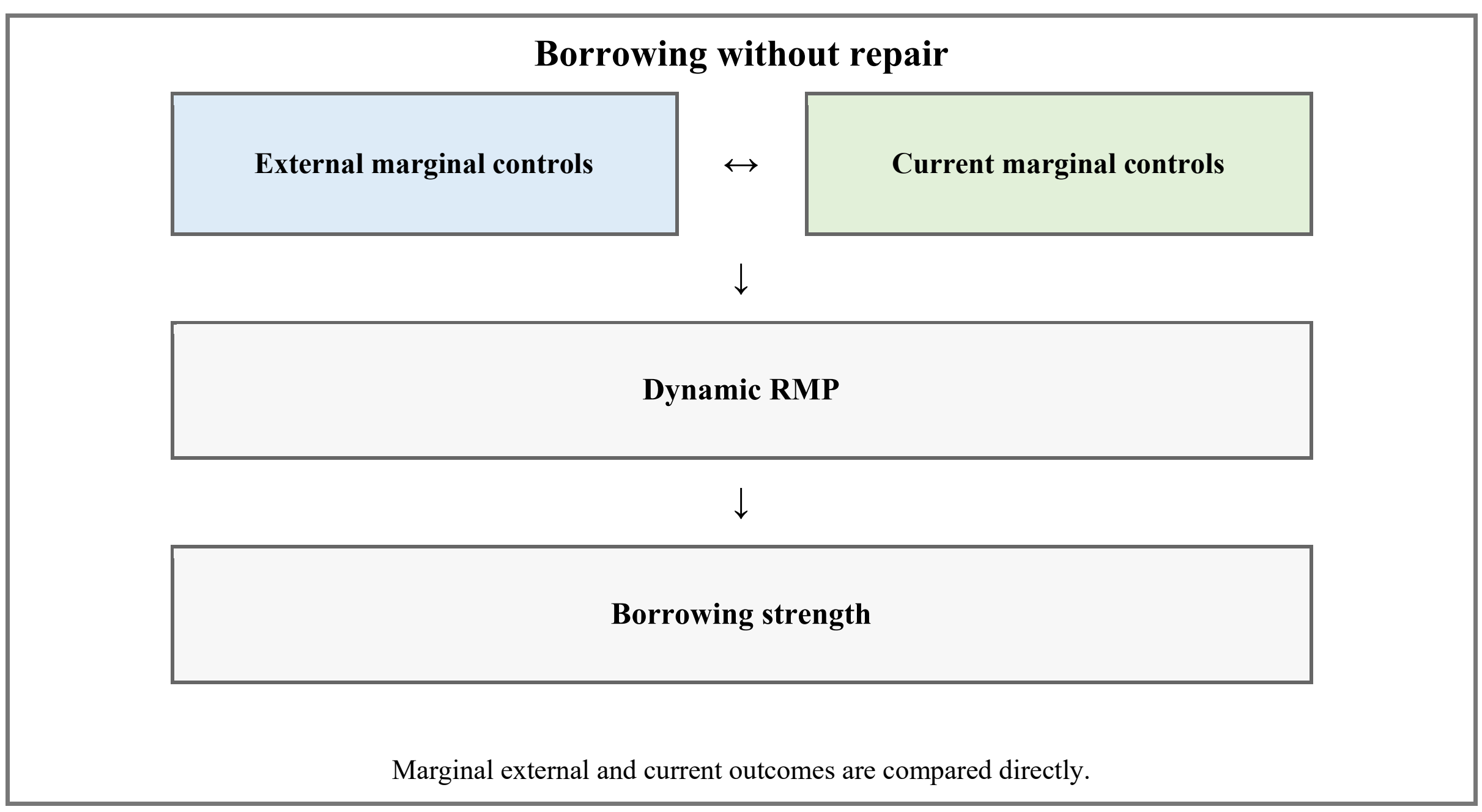

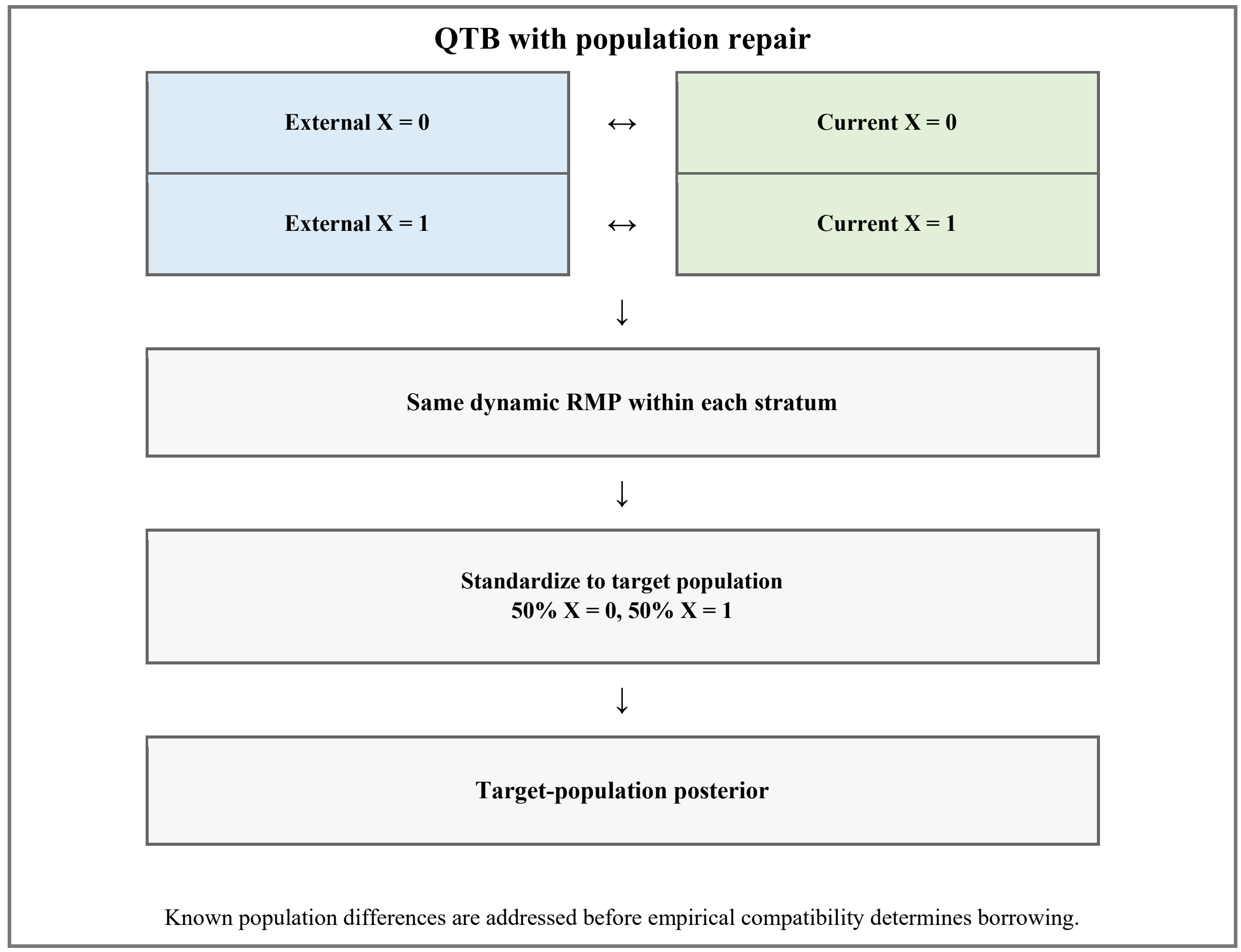

**Align first. Borrow second.**

QTB does not change the RMP updating rule; it changes the external information to which the rule is applied.

Conceptual illustration of where dynamic borrowing operates before and after QTB repair. Without repair, marginal external and current outcomes are compared directly. Under QTB, known population differences are addressed first, the same RMP is applied within aligned strata, and posterior quantities are standardized to the target population.

### Five prespecified scenario families

Five scenario families were prespecified, each targeting a distinct question in the QTB framework. For each family, the information available to QTB for source assessment, the resulting QTB classification, and any required repair analysis were specified before the simulations were run. No simulated outcome realization was used to revise the classification or repair strategy. Some parameters were used only to generate the simulated data and were not treated as information available to QTB; these are identified explicitly below. The same prespecified RMP was used whenever Step 5 allowed borrowing.

**World 1: Fully compatible evidence.** The external controls were generated to match the randomized-trial controls in both population composition and outcome distribution. They had the same covariate distribution,

$$X \sim \text{Bernoulli}(0.5),$$

and the same response probability within each $X$stratum. By construction, no identified material difference required repair or exclusion before borrowing. QTB therefore classified the external data as qualified, and they proceeded unchanged to Step 5, where the prespecified RMP was applied. This world tests whether QTB leaves the borrowing analysis unchanged when the external and current controls are fully compatible.

**World 2: Repairable population shift.** The external controls had a different distribution of $X$ from the target trial: 80% of the external controls had $X = 1$, compared with 50% in the target population. By construction, the external and trial controls had the same response probability within each $X$ stratum; this within-stratum equality was part of the data-generating mechanism and was not used to determine the QTB classification. Because $X$ was measured in both datasets and both strata were represented, the population difference could be addressed through the specified repair analysis. QTB therefore classified the external data as repairable. In Step 5, the same RMP was applied separately within $X = 0$ and $X = 1$, and the resulting stratum-specific control posteriors were then standardized to the target population using equal weights because the target population was 50% $X = 0$ and 50% $X = 1$. This world tests whether a known, measured difference in patient mix should be repaired before borrowing strength is determined, rather than left for marginal outcome disagreement to influence borrowing indirectly.

**World 3: Qualified external data with residual outcome drift.** The external controls had the same covariate distribution as the target trial, and no identified material difference in population, endpoint, measurement, or available prognostic information required repair or exclusion. For data generation only, a logit-scale offset was applied to the external-control outcome model, producing external response probabilities ranging from 0.15 to 0.45. This drift was part of the data-generating mechanism only and was not treated as information available to Steps 1–4 when the source was assessed. The external data were therefore classified as qualified and proceeded unchanged to Step 5, where the same prespecified RMP used in M2 was applied. This world tests whether residual bias can remain after a source is classified as qualified and dynamic borrowing responds to the resulting outcome disagreement by reducing the historical contribution.

**World 4: Known endpoint mismatch.** The external controls were generated from the same underlying response model as the randomized-trial controls, but the recorded external outcomes were subject to a known endpoint-misclassification problem: some patients who truly did not respond were recorded as responders. Specifically,

$$P(Y_E^* = 1 \mid Y_E = 1) = 1,\ P(Y_E^* = 1 \mid Y_E = 0) = q,$$

where

$$q \in \{0.05, 0.10, 0.20\}.$$

Thus, all true responses were recorded correctly, whereas 5%, 10%, or 20% of true nonresponses were incorrectly recorded as responses. The presence of a material endpoint-recording problem was treated as known to QTB, but the patient-level information needed to reconstruct the target endpoint was assumed unavailable. The value of $q$ was used only to generate the simulated data and was not treated as information available to QTB or to the analysis. QTB therefore classified the external data as not qualified for primary borrowing, and no external information proceeded to Step 5.

M1 and M2 nevertheless used the recorded external outcomes, allowing us to evaluate what happens when external data with a known, nonrepairable endpoint problem are still pooled or dynamically borrowed. This world tests whether a known problem that should preclude primary borrowing can instead be left to Step 5 and adequately managed through dynamic downweighting.

This example does not imply that endpoint misclassification is inherently unrepairable. If the information needed to reconstruct the target endpoint were available, the source could instead be classified as repairable and corrected before Step 5.

**World 5: Same outcome distribution, different QTB decisions.** We constructed two external-control settings that were indistinguishable to an outcome-based borrowing model. In both settings, the recorded external response probability was 0.37. In the first setting, this probability arose from residual outcome drift. That drift was part of the data-generating mechanism only and was not treated as information available to Steps 1–4; no identified material difference therefore required repair or exclusion, and QTB classified the external data as qualified. In the second setting, the true external response probability was 0.30, but 10% of true nonresponses were incorrectly recorded as responses, producing the same recorded response probability of 0.37. Because this endpoint-recording problem was known and the information needed to reconstruct the target endpoint was unavailable, QTB classified the external data as not qualified for primary borrowing.

For a given external sample size $n_E$, the recorded number of responses therefore had the same distribution in both settings:

$$Y_E^* \sim \text{Binomial}(n_E, 0.37).$$

Because the two settings generated the same distribution of the observed data used by M2, its borrowing behavior had the same sampling distribution in the two settings. QTB made different pre-borrowing decisions because Steps 1–4 also used prespecified information about endpoint definition and recording that was not contained in the observed response values. This world tests whether observed outcome data alone are sufficient to determine the appropriate pre-borrowing action for external information.

**Calibration, simulation, and operating characteristics**

The same success criterion and cutoff were used for all four analyses. A trial was declared successful when

$$P(\Delta_R > 0 \mid D) > c^*.$$

The cutoff $c^*$was calibrated once using the RCT-only analysis and was then held fixed for M0–M3. Under the prespecified null

$$p_T = p_C = 0.30,$$

exact enumeration of all

$$51 \times 26 = 1{,}326$$

possible pairs of treatment and control response counts gave

$$c^* = 0.972,$$

corresponding to an exact RCT-only Type I error of 0.02254. This was the least stringent attainable cutoff satisfying the prespecified one-sided 0.025 Type I-error target. This calibration was pointwise at the null setting $p_T = p_C = 0.30$and should not be interpreted as uniform Type I-error control over all possible null configurations.

The final simulation used 10,000 replications for each scenario and analysis method. We evaluated posterior-mean bias, RMSE, 95% credible-interval coverage and mean width, Type I error under the null, power under the alternative, posterior historical-component weight where applicable, and relative MSE efficiency,

$$ER_j = \frac{\mathrm{MSE}(M0)}{\mathrm{MSE}(M_j)},\ j \in \{1, 2, 3\}.$$

Values of

$$ER_j > 1$$

indicate lower MSE than the RCT-only analysis. These operating characteristics were considered jointly rather than relying on Type I error or power alone to judge performance.

For the World 2 repair analysis, $K = 5000$posterior draws were used per replication to construct the standardized target-population posterior. This value was selected through the numerical-stability assessment described in the Supplementary Methods. Additional numerical validation, exact checks, random-number control, and other simulation quality-control procedures are also described there. All reported results came from the final simulation run after the design and code had been frozen.

## Results

Results are presented in the order of the scientific questions rather than by World number. We begin with the central question: can observed outcome data alone determine the appropriate pre-borrowing action for external information (World 5)? We then examine what happens when external data with a known, nonrepairable endpoint problem are nevertheless used for borrowing (World 4); whether a measured population difference should be repaired before borrowing strength is determined rather than left to marginal outcome disagreement (World 2); whether residual bias can remain after a source is classified as qualified and dynamic borrowing responds to the remaining outcome disagreement (World 3); and finally whether QTB leaves borrowing unchanged when the external and current data are fully compatible (World 1).

### Observed outcome data alone could not determine the pre-borrowing action

World 5 compared two external-control settings that were indistinguishable to M2 by construction. In both settings, the recorded external response probability was 0.37, so the number of recorded external responses had the same sampling distribution. In one setting, no identified material difference required repair or exclusion, and the external data were classified as qualified. In the other, the recorded endpoint had a known, nonrepairable misclassification problem, and the external data were not qualified for primary borrowing. Because the two settings generated the same distribution of the data used by M2, its operating characteristics were essentially the same in the two settings (Figure 2). The largest standardized between-setting difference across the audited metrics was $|\boldsymbol{Z}| = \mathbf{2.4332}$, consistent with Monte Carlo variation; this comparison was a simulation check rather than a formal equivalence test.

QTB made different pre-borrowing decisions because Steps 1–4 incorporated prespecified scientific information that was not contained in the observed response values. In the qualified setting, the external information proceeded unchanged to Step 5, so M3 = M2. In the setting with the known endpoint error, no external information proceeded to Step 5, so M3 = M0. M2 received the same distribution of observed data in the two settings and therefore could not distinguish why the two settings required different pre-borrowing actions. That distinction came from information about how the endpoint had been defined and recorded. An outcome-based borrowing method cannot distinguish scientific settings that generate the same distribution of the data on which the borrowing rule operates.

### Dynamic downweighting reduced but did not eliminate distortion from a known endpoint problem: World 4

World 4 examined what happened when external data with a known, nonrepairable endpoint problem were nevertheless used for borrowing. QTB classified the source as not qualified for primary borrowing, so no external information proceeded to Step 5 and M3 was identical to the trial-only analysis M0 (Figure 3). The distortion was greatest when the endpoint error was largest ($q = 0.20$)and the external sample size was $n_E = 250$. Under the null, full pooling (M1) produced a bias of $-12.1$percentage points, and coverage of the nominal 95% credible interval fell to 60.8%, even though Type I error was only 0.02%. The RMP (M2) substantially reduced the distortion, but did not eliminate it: bias was $-4.8$percentage points, 95% coverage was 91.8%, and Type I error was 1.0%. In contrast, M3 reproduced the trial-only results: bias was $-0.6$percentage points, coverage was 95.3%, and Type I error was 2.27%.

Because some true nonresponses were recorded as responses, the external control response probability was biased upward. This pulled the estimated control response probability upward and the estimated treatment effect downward, making M1 and M2 less likely to declare the treatment effective and explaining their very low Type I error. Low Type I error therefore did not imply reliable inference: the treatment effect remained biased, and the credible intervals failed to cover the true value too often. The QTB decision to exclude the external information was based on the known endpoint-recording problem and the absence of information needed for repair, not on the operating characteristics observed in the simulation.

**Repair addressed a measured population difference before borrowing strength was determined: World 2**

World 2 examined a measured population difference that could be addressed through the specified repair analysis. The external controls had a different mix of $X = 0$and $X = 1$patients from the target trial, while within each $X$stratum the external and trial controls had the same response probability. M2 applied the RMP directly to the marginal external-control data without first accounting for this known population difference. Under M3, the QTB repair instead required the same RMP to be applied separately within $X = 0$and $X = 1$, after which the stratum-specific control posteriors were standardized to the target trial population.

The advantage of repair became clearer as the external sample grew. At $n_E = 250$, null absolute bias decreased from 1.94 percentage points with M2 to 0.80 percentage points with M3, while RMSE decreased from 0.0832 to 0.0814 and coverage increased from 96.0% to 96.8% (Table 2). Under the alternative, absolute bias decreased from 2.64 to 1.46 percentage points. At $n_E = 25$, repair did not reduce absolute bias, although RMSE and coverage were somewhat better.

Repair did not improve every operating characteristic at every sample size, nor was that its purpose. The known, measured difference in patient mix was addressed by changing the information and comparison structure passed to Step 5 before outcome agreement was used to determine borrowing strength. The alternative was to apply dynamic borrowing to the unadjusted marginal comparison and allow the population difference to influence borrowing only indirectly through the resulting outcome disagreement.

**Residual bias could remain after qualification despite dynamic downweighting: World 3**

World 3 examined a different problem from Worlds 2 and 4. Steps 1–4 identified no material population, endpoint, measurement, or other available-data difference requiring repair or exclusion. The external data therefore proceeded unchanged to Step 5, so M3 used the same RMP as M2 and the two analyses were identical (Figure 4). For data generation, the external response probability was allowed to drift away from the current-control response probability. This created residual outcome disagreement of the kind that Step 5 is intended to address through dynamic borrowing.

The RMP did respond to this disagreement. When the external response probability was 0.15, increasing $n_E$from 50 to 250 reduced the mean posterior historical-component weight from 0.5005 to 0.4495. However, the lower posterior weight did not imply less overall influence from the external information, because the historical component became more informative as the external sample size increased. Type I error increased from 6.62% to 10.02%. At $n_E = 250$, bias was 3.85 percentage points and 95% coverage fell to 89.2%. Independent exact enumeration gave a Type I error of 9.45% for the same setting, close to the simulation result and confirming that the inflation was not a coding or Monte Carlo artifact. Full pooling was much more severely affected, with Type I error of 62.7% and coverage of only 39.0%.

When the external response probability was below the current-control value of 0.30, the estimated control response probability was pulled downward, making false-positive treatment conclusions more likely. When the external response probability was above 0.30, the bias went in the opposite direction and null rejection probabilities fell to 0.8%–1.3%. Thus, Step 5 responded to residual empirical incompatibility by reducing the posterior weight on the historical component, but dynamic downweighting did not guarantee control of residual bias or Type I error. Classification as qualified determined that the external information could proceed unchanged to Step 5; it did not certify exchangeability or guarantee that the remaining source difference would be fully neutralized.

**QTB reduced to the usual borrowing analysis when the external data were fully compatible: World 1**

World 1 examined the case in which Steps 1–4 identified no material difference requiring repair or exclusion. The external data therefore proceeded unchanged to Step 5. Because $D_E^* = D_E$, M3 used exactly the same RMP as M2, and the two analyses were identical.

Compared with M0, M2/M3 had lower null MSE, with MSE-efficiency ratios increasing from 1.260 to 1.711 as $n_E$increased, and narrower mean 95% credible intervals, with null widths decreasing from 0.387 to 0.347 versus approximately 0.417 for M0 (Table 2). Under the alternative, the rejection probability for M2/M3 increased from 46.6% at $n_E = 25$to 58.7% at $n_E = 250$. These gains were accompanied by lower MSE, slightly above-nominal coverage, and conservative null rejection.

When no identified material difference required repair or exclusion, the external information passed unchanged to Step 5 and QTB reduced exactly to the usual RMP analysis. The efficiency gains from borrowing were therefore preserved.

**Discussion**

The main message is not that Bayesian borrowing should be avoided, nor that empirical agreement between external and current outcomes is unimportant. The issue is when that agreement should enter the borrowing process and what external information it should be used to evaluate. Existing approaches already recognize the need to assess source relevance, population comparability, endpoint alignment, and potential bias before borrowing. QTB makes that separation explicit and operational: known material differences are first used to determine whether the external data should be used directly, repaired before use, or excluded from primary borrowing. Only after those decisions have been made does the selected Bayesian borrowing method use the remaining agreement or disagreement with the current data to determine how strongly the external information contributes. In QTB, dynamic borrowing therefore operates on residual empirical compatibility rather than on empirical compatibility before known differences have been addressed.

World 5 shows why this separation is needed. Two external-control settings gave the borrowing model the same distribution of observed responses, yet one had no known problem requiring repair or exclusion and the other had a known endpoint error. An outcome-based borrowing method could not distinguish them because the information needed to make that distinction was not contained in the outcome data it used. World 4 shows why this matters. When data with a known endpoint problem were nevertheless borrowed, the RMP reduced their influence and substantially attenuated the resulting distortion, but bias and undercoverage remained. Very low Type I error did not mean that the inference was reliable. The fact that dynamic borrowing reduced the influence of the problematic external data does not mean that those data were appropriate for primary borrowing. Reducing the impact of an external source and deciding whether that source should enter borrowing are different questions.

Not every identified difference should lead to exclusion. World 2 illustrates the middle path built into QTB: a source may be unsuitable for direct borrowing but still usable after a prespecified repair. The external controls differed from the target trial in patient mix, but the relevant difference was measured and could be addressed before borrowing. The purpose of repair was not to guarantee improvement in every operating characteristic, nor did it do so at every sample size. Rather, repair changed the external information presented to the borrowing model so that the known population difference was addressed before empirical compatibility was used to determine borrowing. QTB therefore treats repair as a distinct action between direct borrowing and exclusion: when a material difference is known and adequately addressable, the appropriate response is to repair the information first, not simply to borrow less from it or discard it altogether.

Qualification does not eliminate the need for dynamic borrowing. World 3 shows why Step 5 remains necessary even after an external source has passed the pre-borrowing assessment in Steps 1–4. No identified material difference required repair or exclusion, yet residual outcome drift was still present in the data-generating mechanism. The RMP, operating as Step 5 of QTB, responded to that residual disagreement by reducing the posterior weight on the historical component. Bias, undercoverage, and Type I-error inflation could nevertheless remain. Thus, qualification and dynamic borrowing are sequential parts of the same framework: Steps 1–4 address identifiable differences before borrowing, and Step 5 uses residual empirical compatibility to determine how much of the remaining external information should contribute. The latter is an additional safeguard, not a guarantee that unrecognized source differences or residual bias have been eliminated.

World 1 shows the complementary case in which Steps 1–4 require no modification of the external information. When the external and current controls were fully compatible by construction, the external data passed unchanged to Step 5, so $D_E^* = D_E$. The resulting QTB analysis was therefore exactly the usual RMP analysis: M3 and M2 were identical, and the efficiency gains from borrowing were preserved. Thus, conventional dynamic borrowing is naturally retained within QTB when no repair or exclusion is required. QTB changes the borrowing analysis only when the pre-borrowing assessment identifies a reason to change the information that reaches Step 5.

The RMP used in this study is one implementation of Step 5, not a defining component of QTB. QTB is a borrowing framework rather than a new Bayesian prior, and other dynamic borrowing approaches—including robust MAP priors, commensurate priors, power priors, and explicit bias models—could be used in the same role. Different methods may quantify empirical compatibility and translate it into borrowing in different ways. What QTB specifies is the sequence in which that compatibility is allowed to matter: known material differences are addressed through Steps 1–4, and the selected borrowing method is then applied to the external information that remains. Residual empirical compatibility therefore refers to the stage and information on which borrowing operates, not to a new universal compatibility statistic.

Recent Bayesian borrowing methods also go beyond simple outcome-adaptive downweighting by modeling source bias explicitly. Potential-bias models, for example, introduce bias parameters between historical and current controls and use shrinkage priors to regulate how much historical information is borrowed. [13] More recently, B-CALM jointly aligns randomized and observational data and uses explicit baseline- and comparative-bias functions to limit the contribution of observational evidence to an RCT-defined estimand. [9] These developments are complementary to QTB rather than alternatives to it. QTB does not claim that residual source bias must be handled only through empirical outcome conflict. Instead, its contribution is to determine what should happen before the downstream borrowing model is applied: an identified material difference may require repair, may preclude primary borrowing, or may leave residual uncertainty that an explicit bias model can represent. After that routing decision, Step 5 may use either outcome-adaptive borrowing or an explicit bias model to regulate the contribution of the remaining external information. These combinations were not evaluated here and warrant further study.

Taken together, the five scenario families clarify where the contribution of QTB lies. QTB does not introduce new principles for defining estimands, assessing external-data quality, or addressing measured differences; those ideas are well established. Its contribution is to organize those assessments into a prespecified decision sequence with explicit consequences for borrowing: external information may proceed unchanged, may require repair before proceeding, or may be excluded from primary borrowing. Dynamic borrowing then operates as the final step on the residual empirical compatibility of the information that remains. The simulations were designed to examine this decision architecture, rather than to suggest that careful assessment of external data is itself new.

A practical implication of QTB is that a Bayesian borrowing analysis should document more than the prior and the mechanism used to modulate borrowing. Investigators should make explicit the target estimand, the external-data features considered in qualification, any material differences identified, the prespecified action taken for each difference, and the form of external information ultimately passed to Step 5. This documentation makes the path from scientific assessment to borrowing transparent and auditable: readers can see not only how much external information was borrowed, but also why that information was allowed to enter borrowing in that form.

The worked oncology examples sharpen an additional point: repair is alignment, not automatic discounting. A valid repair can make target-aligned external information more or less empirically compatible with the current data, so it can lead to stronger or weaker borrowing in Step 5. Likewise, the same apparent mismatch can be repairable in one source and not qualified in another when the information needed for repair is present in one source but absent in the other. The relevant question is therefore not simply what type of difference is present, but whether that difference can be addressed for the prespecified target using information that is actually available.

**Limitations and future directions**

Several limitations define what this study establishes and what remains to be developed. First, QTB provides a decision architecture rather than a validated operational rule for classification. In the simulations, the information available for source assessment and the resulting classifications were prespecified. The study therefore does not establish how consistently different investigators would classify the same real-world source, what threshold should define a material or adequately repairable difference in a given application, or how sensitive inference is to classification error. Developing structured assessment criteria, documentation standards, and sensitivity analyses for uncertain classifications is an important next step. Second, the study validates only selected implementations of the framework. World 2 used one repair based on stratification and standardization, and Step 5 was evaluated using one dynamic RMP. Alternative repair strategies, explicit bias models, and other Bayesian borrowing engines may behave differently, and uncertainty introduced by the repair itself was not studied.

A more fundamental limitation is that QTB can act only on differences that are identified with the information available before borrowing. Unrecognized source differences may remain after qualification and repair, and dynamic borrowing in Step 5 may reduce their influence without fully removing the resulting bias. QTB therefore does not certify exchangeability or guarantee Type I-error control; operating characteristics remain design- and model-specific and require their own calibration and sensitivity assessment. Finally, the simulation setting was intentionally simple, with one randomized trial, one external-control dataset, one binary endpoint, and one binary prognostic covariate. The continuous- and time-to-event examples in the Supplementary Material illustrate implementation but do not provide simulation-based validation of operating characteristics. Extensions to multiple external sources, richer covariate structures, complex time-to-event outcomes, platform trials, and other settings will require corresponding qualification, repair, and validation strategies.

**QTB places dynamic Bayesian borrowing within a prespecified sequence of scientific and statistical decisions.** Before empirical compatibility is used to determine the amount of borrowing, identified material differences are assessed and the external information is either passed forward directly, repaired before use, or excluded from primary borrowing. **Dynamic borrowing then operates on the residual empirical compatibility of the information that remains.** This sequence does not eliminate uncertainty about external-data validity or residual bias, but it makes explicit which problems are addressed before borrowing and which are left for the borrowing model to manage.

## Funding

Dr. Pan's research was supported by the Comprehensive Cancer Center at St. Jude Children's Research Hospital and American Lebanese Syrian Associated Charities.

# Qualify–Then–Borrow

Five-step QTB framework for Bayesian borrowing

**Steps 1–4 determine what external information reaches Step 5 and in what form; Step 5 determines borrowing strength.**

## STEPS 1–4: PRE-BORROWING ASSESSMENT AND ACTION

**1 Define the target**
Target estimand and population
Treatment strategy / condition
Outcome and measurement process
Time origin, follow-up, intercurrent events

→

**2 Assess the external data**
Estimand and control-strategy alignment
Eligibility, time zero, and follow-up
Endpoint definition and ascertainment
Target-population support and overlap
Prognostic covariates / adjustment information
Data quality and provenance

*For each identified material difference: is the information needed for repair available?*
**Observed outcome agreement is not used to resolve Steps 1–4.**

↓

**3 Classify the external source**

| Qualified | Repairable | Not qualified for primary borrowing |
|---|---|---|
| ↓ | ↓ | ↓ |

**4 Use directly, repair before use, or exclude**

| Proceed unchanged to Step 5 | Proceed through specified repair | No external information proceeds to Step 5 |
|---|---|---|
| External information passed to Step 5: $D_E^*$ | | Trial-only primary inference |

↓

## STEP 5: DYNAMIC BAYESIAN BORROWING

$(D_E^*, D_C)$ → **Selected Bayesian borrowing method** Outcome-adaptive or bias-aware implementation → **Borrowing strength** How strongly external information contributes

**Residual empirical compatibility** describes the agreement or source uncertainty that remains after identified material differences have been addressed; it is not a universal compatibility statistic.

**Steps 1–4 determine what reaches Step 5; Step 5 determines borrowing strength.**

IMPORTANT LIMITATION

**Residual bias may still remain.**

Qualification does not certify exchangeability, and Step 5 does not guarantee elimination of residual bias or Type I-error inflation.

**Figure 1. Five-step Qualify-Then-Borrow framework for Bayesian borrowing.** Step 1 defines the prespecified target. Step 2 assesses the external data relative to that target using scientific and data information without using observed outcome agreement to resolve the assessment. Step 3 classifies the source as qualified, repairable, or not qualified for primary borrowing. Step 4 determines what external information, if any, proceeds to Step 5 and in what form: unchanged, through a specified repair, or not at all. The resulting external information is denoted $D_E^*$. Step 5 is the dynamic Bayesian borrowing step: the selected borrowing method operates on $D_E^*$ together with the current data and determines borrowing strength. QTB refers to the agreement or source uncertainty that remains after identified material differences have been addressed as residual empirical compatibility; this is a conceptual stage of borrowing rather than a new universal compatibility statistic. Qualification does not establish exchangeability, and residual bias may remain after Step 5.

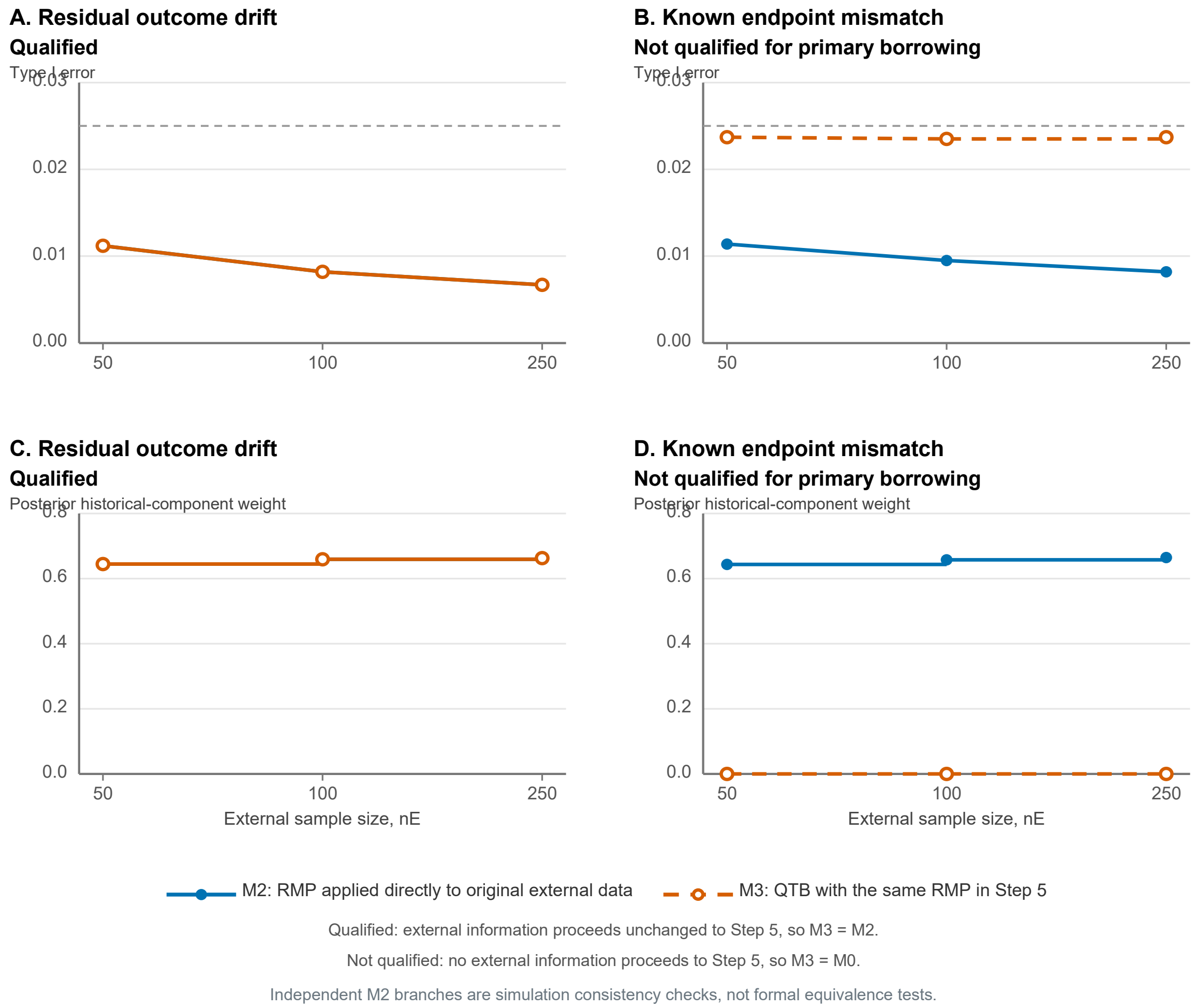

**Figure 2. Same observed outcome distribution, different QTB decisions.** World 5 compares two external-control settings with the same recorded external response probability, 0.37, and therefore the same sampling distribution of the observed data used by M2. In the residual-drift setting, Steps 1–4 identified no material difference requiring repair or exclusion, so the external information proceeded unchanged to Step 5 and M3 = M2. In the known endpoint-mismatch setting, the source was classified as not qualified for primary borrowing, so no external information proceeded to Step 5 and M3 = M0. Because M2 received the same observed-data distribution in the two settings, its operating characteristics were essentially the same. The different M3 results reflect different pre-borrowing decisions based on prespecified scientific information not contained in the observed response values. Differences between independently simulated M2 branches serve as simulation consistency checks rather than formal equivalence tests.

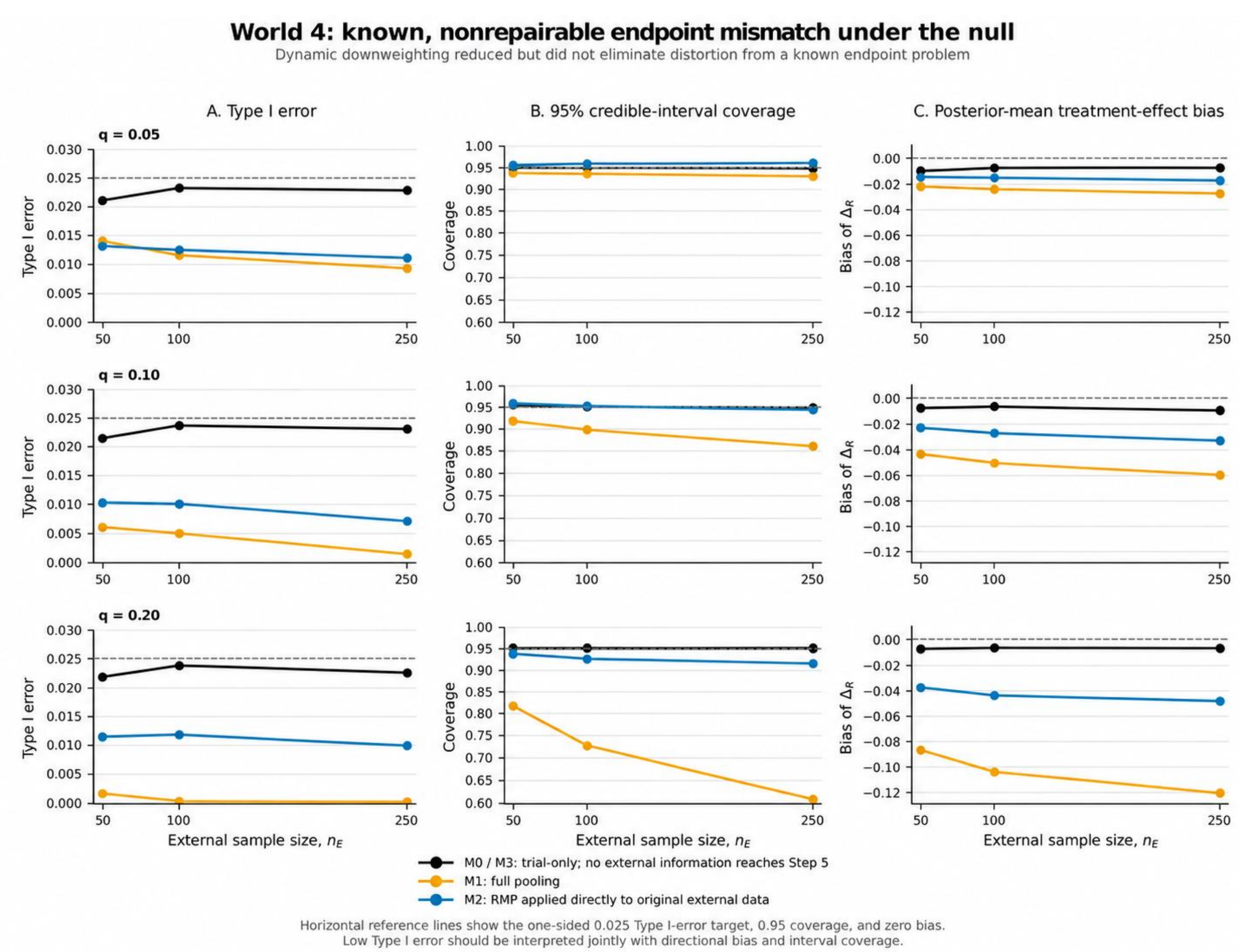

**Figure 3. Dynamic downweighting reduced but did not eliminate distortion from a known endpoint problem.** World 4 null results are shown across external sample sizes n_E = 50, 100, and 250 and false-positive endpoint-misclassification probabilities q = 0.05, 0.10, and 0.20. Panel A shows Type I error, Panel B shows 95% credible-interval coverage, and Panel C shows posterior-mean treatment-effect bias. M0 is the trial-only analysis. M1 fully pools the external and concurrent controls. M2 applies the RMP directly to the original external data without QTB Steps 1–4. Under QTB, the known endpoint mismatch is not repairable with the available information, so no external information proceeds to Step 5 and M3 is identical to M0. Horizontal reference lines indicate the one-sided 0.025 Type I-error target, 0.95 coverage, and zero bias. As endpoint misclassification and external sample size increase, M1 and M2 can become increasingly conservative while simultaneously exhibiting greater treatment-effect bias and poorer interval coverage. Low rejection probability therefore should not be interpreted as evidence of reliable inference in isolation.

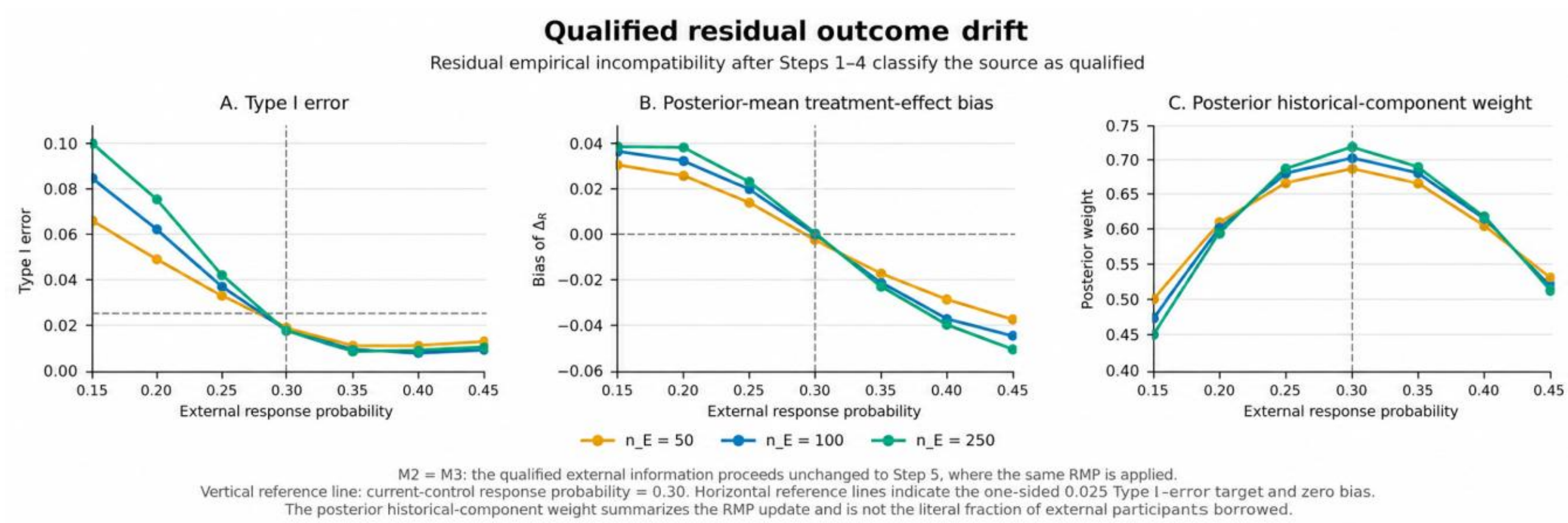

**Figure 4. Residual bias could remain after qualification despite dynamic downweighting.** World 3 null results are shown across external response probabilities from 0.15 to 0.45 and external sample sizes n_E = 50, 100, and 250, with the current-control response probability fixed at 0.30. Steps 1–4 identified no material difference requiring repair or exclusion, so the external information proceeded unchanged to Step 5 and M3 was identical to M2. Panel A shows Type I error, Panel B shows posterior-mean treatment-effect bias, and Panel C shows the posterior historical-component weight from the RMP. The vertical reference line marks the current-control response probability of 0.30; horizontal reference lines indicate the one-sided 0.025 Type I-error target and zero bias. The RMP changed the posterior historical-component weight as residual outcome disagreement increased, but dynamic downweighting did not guarantee control of residual bias or Type I error. The posterior historical-component weight summarizes the RMP update and should not be interpreted as the literal fraction of external participants borrowed.

# QTB Main-Text Tables

**Table 1. QTB classifications, pre-borrowing actions, and consequences for Step 5**

| Classification | Meaning | Step 4: what happens | Step 5: borrowing | Illustrative example |
|---|---|---|---|---|
| **Qualified** | No identified material difference requires repair or exclusion before Step 5. | External information proceeds unchanged to Step 5. | The selected Bayesian borrowing method operates on the external data as observed and determines how strongly they contribute. | Same endpoint definition and time origin, with adequate support for the target population and no identified material difference requiring pre-borrowing action. |
| **Repairable** | One or more identified material differences require action before Step 5, but each can be adequately addressed using the available data and a scientifically justified repair. | External information proceeds only through the specified repair analysis. | The same borrowing method operates on the repaired or target-aligned external information and determines how strongly it contributes. | The external controls contain a different proportion of patients with a measured high-risk factor; the factor is available in both datasets with adequate overlap, allowing stratification, weighting, or standardization to the target population. |
| **Not qualified for primary borrowing** | At least one identified material difference cannot be adequately addressed with the available data. | No external information proceeds to Step 5 for the primary analysis. | In this demonstration, the primary analysis reduces to randomized-trial-only inference. | The target is enrollment- or leukapheresis-based CAR-T event-free survival, but the external dataset contains only infused patients and the earlier treatment course or omitted non-infused patients cannot be recovered. |

**Note:** QTB classification is based on information about the target and external source that is available before observed outcome agreement is used to determine borrowing strength. For each identified material difference, QTB asks whether the information needed to address it is available and what action is required before Step 5. Classification determines the pre-borrowing action; it does not establish that the external and current data are exchangeable. In this demonstration, sources not qualified for primary borrowing contribute no external information to Step 5, and inference uses the randomized-trial data alone.

**Table 2. Key operating characteristics for fully compatible evidence and a repairable population shift**

| Setting | $n_E$ | Method | Null bias | Null RMSE | Null coverage | Null width | Type I error | Alt bias | Alt RMSE | Alt coverage | Power | Null ER |
|---|---|---|---|---|---|---|---|---|---|---|---|---|
| World 1 | 25 | M0 | -0.0065 | 0.1055 | 0.9539 | 0.4166 | 0.0215 | -0.0130 | 0.1099 | 0.9530 | 0.4123 | 1.0000 |
| World 1 | 25 | M2 = M3 (qualified) | -0.0022 | 0.0940 | 0.9617 | 0.3871 | 0.0199 | -0.0085 | 0.0991 | 0.9601 | 0.4661 | 1.2604 |
| World 1 | 50 | M0 | -0.0078 | 0.1047 | 0.9536 | 0.4167 | 0.0201 | -0.0160 | 0.1099 | 0.9493 | 0.4003 | 1.0000 |
| World 1 | 50 | M2 = M3 (qualified) | -0.0015 | 0.0883 | 0.9652 | 0.3720 | 0.0174 | -0.0095 | 0.0934 | 0.9609 | 0.4994 | 1.4075 |
| World 1 | 100 | M0 | -0.0082 | 0.1058 | 0.9509 | 0.4171 | 0.0240 | -0.0166 | 0.1104 | 0.9491 | 0.3970 | 1.0000 |
| World 1 | 100 | M2 = M3 (qualified) | -0.0006 | 0.0844 | 0.9636 | 0.3595 | 0.0196 | -0.0091 | 0.0897 | 0.9654 | 0.5379 | 1.5709 |
| World 1 | 250 | M0 | -0.0085 | 0.1047 | 0.9563 | 0.4168 | 0.0191 | -0.0152 | 0.1100 | 0.9493 | 0.4055 | 1.0000 |
| World 1 | 250 | M2 = M3 (qualified) | 0.0002 | 0.0801 | 0.9688 | 0.3469 | 0.0168 | -0.0063 | 0.0852 | 0.9685 | 0.5873 | 1.7114 |
| World 2 | 25 | M2 (direct RMP) | -0.0115 | 0.0947 | 0.9607 | 0.3892 | 0.0161 | -0.0197 | 0.1010 | 0.9550 | 0.4208 | 1.2186 |
| World 2 | 25 | M3 (QTB repair) | -0.0130 | 0.0931 | 0.9622 | 0.3828 | 0.0139 | -0.0214 | 0.0991 | 0.9571 | 0.4215 | 1.2609 |
| World 2 | 50 | M2 (direct RMP) | -0.0155 | 0.0911 | 0.9598 | 0.3748 | 0.0125 | -0.0213 | 0.0971 | 0.9556 | 0.4468 | 1.3449 |
| World 2 | 50 | M3 (QTB repair) | -0.0127 | 0.0898 | 0.9607 | 0.3732 | 0.0132 | -0.0187 | 0.0956 | 0.9594 | 0.4525 | 1.3848 |
| World 2 | 100 | M2 (direct RMP) | -0.0175 | 0.0874 | 0.9597 | 0.3629 | 0.0127 | -0.0243 | 0.0930 | 0.9575 | 0.4665 | 1.4605 |
| World 2 | 100 | M3 (QTB repair) | -0.0105 | 0.0859 | 0.9635 | 0.3635 | 0.0145 | -0.0174 | 0.0910 | 0.9628 | 0.4805 | 1.5126 |
| World 2 | 250 | M2 (direct RMP) | -0.0194 | 0.0832 | 0.9604 | 0.3509 | 0.0090 | -0.0264 | 0.0906 | 0.9561 | 0.4935 | 1.6125 |
| World 2 | 250 | M3 (QTB repair) | -0.0080 | 0.0814 | 0.9684 | 0.3526 | 0.0136 | -0.0146 | 0.0870 | 0.9637 | 0.5168 | 1.6866 |

**Note:** Results are based on 10,000 replications per scenario and analysis method. World 1 represents fully compatible evidence; World 2 represents the repairable population-shift setting. Alt = alternative. ER = MSE(M0)/MSE(method); values greater than 1 indicate lower MSE than the trial-only analysis. In World 1, no identified material difference required repair or exclusion, so the external information proceeded unchanged to Step 5 and M2 and M3 were identical; they are therefore shown once. In World 2, M2 applies the RMP directly to the marginal external data, whereas M3 applies the same RMP separately within X = 0 and X = 1 and then standardizes the stratum-specific posterior quantities to the target trial's 50%/50% covariate distribution. Complete results for all methods, including M1, are provided in the Supplementary Results.

# Supplementary Methods, Worked Examples, and Results

*Qualify-Then-Borrow: A Five-Step Framework for Bayesian Borrowing Beyond Outcome Agreement*

## Supplementary Methods

### S1. Statistical setting and notation

The demonstration contains one randomized trial with an experimental-treatment arm and a concurrent-control arm, together with one external-control dataset. The endpoint Y is binary, with Y=1 denoting response, and X is a binary prognostic baseline covariate. The target population is the randomized-trial population. Let $p_T$ and $p_C$ denote its marginal treatment and control response probabilities. The target estimand is the marginal risk difference

$$\Delta_R = p_T - p_C$$

External data may inform only $p_C$. They do not change the target population, outcome, treatment contrast, or estimand. Trial sizes are $n_T=50$ and $n_C=25$. The external sample size $n_E$ takes values 25, 50, 100, and 250, with the exact grid depending on the scenario family.

### S2. Data-generating model

In the randomized-trial target population,

$$X \sim \text{Bernoulli}(0.5)$$

For current controls, X=1 doubles the odds of response relative to X=0. The logistic model is

$$\text{logit}\Pr(Y = 1 \mid A = 0, S = 1, X) = \beta_0 + \beta_X X, \qquad \beta_X = \log 2$$

The intercept $\beta_0$ is chosen so that the marginal current-control response probability is $p_C=0.30$; numerically $\beta_0 \approx -1.2180165702$. Under the alternative treatment condition,

$$\text{logit}\Pr(Y = 1 \mid A = 1, S = 1, X) = \beta_0 + \beta_X X + \beta_A$$

with $\beta_A$ chosen so that the marginal treatment response probability is $p_T=0.50$; numerically $\beta_A \approx 0.8714429800$. Thus $\Delta_R=0.20$ under the alternative. Under the null, $\beta_A=0$, so $p_T=p_C=0.30$ and $\Delta_R=0$.

### S3. Five-step QTB framework

QTB is not a new Bayesian prior or borrowing model. It is a five-step decision framework. Steps 1–4 determine what external information, if any, proceeds to the dynamic borrowing step and in what form. These decisions are based on the prespecified target and source information available before observed outcome agreement is used to determine borrowing strength. Step 5 is the dynamic borrowing step: the selected Bayesian borrowing method operates on the information passed forward from Step 4 and determines how strongly it contributes.

The five steps are: (1) define the target; (2) assess the external data; (3) classify the external source as qualified, repairable, or not qualified for primary borrowing; (4) use the external information directly, repair it before use, or exclude it; and (5) determine how much information to borrow.

Let S_E denote the QTB classification: qualified (Q), repairable (R), or not qualified for primary borrowing (N). Let D_E denote the original external data and T the prespecified target. At the framework level, the external information passed to Step 5 is denoted D_E*:

$$D_E^\star = \begin{cases} D_E, & S_E = Q, \\ \mathcal{R}(D_E;\ \mathrm{T}), & S_E = R, \\ \varnothing, & S_E = N. \end{cases}$$

Here, R(D_E; T) denotes the specified repair or target-aligned representation through which external information is allowed to enter Step 5. In World 2, for example, the repair defines a stratum-specific borrowing and target-population standardization analysis. If D_E*=∅, no external information proceeds to Step 5 for the primary analysis.

After Steps 1–4, the external and current data may still agree or disagree. QTB calls the agreement or disagreement that remains after identified material differences have been addressed residual empirical compatibility.

Residual empirical compatibility is not a new universal compatibility statistic. It defines the stage and the external information on which the selected Bayesian borrowing method operates. Different Step-5 methods may translate residual outcome agreement, commensurability, or residual source uncertainty into borrowing in different ways.

In this study, Step 5 is implemented with the robust mixture prior (RMP) described next. The RMP is therefore one implementation of the final QTB step, not a separate method applied after QTB.

## S4. Robust mixture prior used in Step 5

For a qualified external source, D_E*=D_E, so Step 5 applies the robust mixture prior (RMP) to the external data as observed. The historical component is constructed from the external controls, while the vague Beta(1,1) component represents little prior information about the control response probability. For y_E responses among n_E external controls,

$$\pi_H(p_C) = \mathrm{Beta}(1 + y_E,\ 1 + n_E - y_E)$$

The vague component and initial historical-component weight are

$$\pi_V(p_C) = \mathrm{Beta}(1,1), \qquad w_0 = 0.50$$

$$\pi(p_C \mid D_E) = w_0 \pi_H(p_C) + (1 - w_0)\pi_V(p_C)$$

After y_C responses among n_C concurrent controls, the current data update both Beta components and the relative weight assigned to them. Let a_H=1+y_E and b_H=1+n_E−y_E. The predictive probabilities of the observed concurrent-control data under the historical and vague components are

$$L_H = \binom{n_C}{y_C} \frac{B(a_H + y_C,\ b_H + n_C - y_C)}{B(a_H, b_H)}$$

$$L_V = \binom{n_C}{y_C} \frac{B(1 + y_C,\ 1 + n_C - y_C)}{B(1,1)}$$

where B(a,b) is the beta function. The posterior historical-component weight is

$$w_{\text{post}} = \frac{w_0 L_H}{w_0 L_H + (1 - w_0) L_V}$$

With $w_0$=0.50, equal predictive support leaves the historical-component weight at 0.50. If the historical component predicts the current controls better than the vague component, w_post increases; if it predicts them worse, w_post decreases. Thus, the RMP determines how strongly the historical information contributes through the relative predictive support provided by the current-control data.

The resulting posterior for the control response probability is

$$p_C \mid D_E, D_C \sim w_{\text{post}} \text{Beta}(a_H + y_C,\, b_H + n_C - y_C) + \left(1 - w_{\text{post}}\right) \text{Beta}(1 + y_C,\, 1 + n_C - y_C)$$

The first posterior component combines the external and current control responses; the second uses the current controls alone. This is the usual RMP update; QTB adds no new compatibility statistic or compatibility test at this stage. In this RMP implementation, the relative predictive support provided by the current data operationalizes residual empirical compatibility after identified material differences have been addressed in Steps 1–4. The posterior mixture weight describes how strongly the historical component contributes, but it is not the literal proportion of external participants borrowed.

## S5. Repairable population shift: stratification and standardization

World 2 uses P_E(X=1)=0.80, while the target trial population has P_R(X=1)=0.50. Within each X stratum, the external and current controls have the same response probability by construction; this within-stratum equality is part of the data-generating mechanism and is not used to determine the QTB classification. Because X is measured in both datasets and both strata are represented, the source is classified as repairable. The repair defines the comparison and standardization structure: external and concurrent controls are separated by X, and the same RMP described in S4 is applied separately within each stratum. Within stratum x∈{0,1}, let n_Ex and y_Ex denote the external sample size and number of responses, and let n_Cx and y_Cx denote the corresponding concurrent-control values.

The posterior historical-component weight in stratum x is

$$w_{\text{post},x} = \frac{w_0 L_{H,x}}{w_0 L_{H,x} + (1 - w_0) L_{V,x}}$$

where L_H,x and L_V,x are the stratum-specific predictive probabilities defined in the same way as in S4.

The control posterior in stratum x is

$$p_{Cx} \mid D \sim w_{\text{post},x} \text{Beta}(1 + y_{Ex} + y_{Cx},\, 1 + n_{Ex} - y_{Ex} + n_{Cx} - y_{Cx}) + \left(1 - w_{\text{post},x}\right) \text{Beta}(1 + y_{Cx},\, 1 + n_{Cx} - y_{Cx})$$

Thus, X=0 and X=1 each have their own RMP update and their own posterior control response probability. External and current controls are compared only within the same covariate stratum; this is the comparison structure specified by the repair.

Let p_C0^(b) and p_C1^(b) denote posterior draw b from the two stratum-specific control distributions. Because the target trial population is 50% X=0 and 50% X=1, the stratum-specific posterior quantities are standardized to the target population as follows:

$$p_C^{R,(b)} = 0.5\, p_{C0}^{(b)} + 0.5\, p_{C1}^{(b)}, \qquad \Delta_R^{(b)} = p_T^{(b)} - p_C^{R,(b)}$$

If a stratum contains no external controls, that stratum uses the concurrent-control-only posterior and receives no external borrowing. If neither dataset contains observations in that stratum, the posterior is Beta(1,1). External information is never transferred between X strata. No propensity-score adjustment is used in this demonstration.

## Algorithm S1. Practical QTB implementation for a binary endpoint

**Purpose.** This algorithm translates the five-step QTB framework into an implementable workflow for the binary-endpoint setting used in this paper. The code does not perform source assessment or classification. It begins only after the target has been defined, the external source has been assessed, the QTB classification has been documented, and any required repair analysis has been specified.

**Inputs.**

- Trial data: treatment responses y_T of n_T and concurrent-control responses y_C of n_C.
- Candidate external data: external responses y_E of n_E, or stratum-specific counts when a repair analysis is prespecified.
- A prespecified target estimand; here, the target-population risk difference

$$\Delta_R = p_T - p_C$$

- The QTB classification—qualified, repairable, or not qualified for primary borrowing—determined and documented before observed outcome agreement is used to determine borrowing strength.
- The specified repair analysis when the source is repairable; in World 2, this is RMP borrowing within X strata followed by standardization to the target population.
- The Step-5 borrowing specification; here, the Beta-Binomial RMP with w0 = 0.50. If a trial decision is required, use the prespecified cutoff c* rather than recalibrating it after seeing the data. The code examples below use the full-precision calibrated cutoff; the main text reports the rounded value c*=0.972.

**Operational steps.**

**1. Define the target.** Specify the estimand and target population before assessing the external source.

**2. Assess the external data.** Relative to the prespecified target, assess estimand and control-strategy alignment; eligibility, time zero, and follow-up; endpoint definition and ascertainment; target-population support and overlap; prognostic covariates needed for adjustment; data quality and provenance; and whether any identified material difference can be adequately addressed with the available information. Observed outcome agreement is not used to resolve this assessment.

**3. Classify the external source.** Classify the source as qualified, repairable, or not qualified for primary borrowing according to what must happen before Step 5. Classification determines the pre-borrowing action; it does not establish exchangeability.

**4. Use directly, repair before use, or exclude.** Qualified: external information proceeds unchanged to Step 5. Repairable: external information proceeds only through the specified repair analysis. Not qualified for primary borrowing: no external information proceeds to Step 5 for the primary analysis; in this demonstration, the control analysis is trial-only.

**5. Determine how much information to borrow.** Apply the selected RMP to the external information D_E* passed forward from Step 4. For a qualified source, D_E*=D_E; for a repairable source, D_E* is the repaired or target-aligned representation; for a source not qualified for primary borrowing, D_E*=∅. The current-control data update the historical and vague components, producing the posterior historical-component weight w_post. This is the RMP-specific response to the residual empirical compatibility that remains after Steps 1–4.

**After the five QTB steps: posterior summaries and sensitivity assessment**

**Form and summarize the treatment-effect posterior.** Combine the trial-only treatment posterior with the Step-5 control posterior. For the repairable World-2 example, obtain stratum-specific control posterior draws and standardize them using the target-population weights before forming Delta_R. Report the posterior mean, 95% credible interval, Pr(Delta_R > 0 | D), and the posterior historical-component weight (or a prespecified summary of stratum-specific weights). If the design uses a success cutoff, compare the posterior probability with the locked c*. The historical-component weight is a model-specific reporting quantity, not evidence that the source is scientifically qualified and not a literal fraction of external participants borrowed.

**Assess residual bias and sensitivity.** Classification as qualified does not establish exchangeability, and Step 5 does not guarantee elimination of residual bias or Type I-error inflation. Prespecified sensitivity analyses or bias-aware borrowing methods remain appropriate when residual source differences are plausible.

**Minimal R implementation.**

The code below is a transparent posterior-computation example for the paper's binary setting. It mirrors the same Beta-Binomial RMP and the three Step-4 routes, but it is not the production simulation code. For simplicity, it uses posterior Monte Carlo for all routes. The status argument must be assigned before the code is run; the code does not perform source assessment or QTB classification. The full-precision cstar shown in the examples corresponds to the rounded cutoff c*=0.972 reported in the main text.

```
rmp_beta_binom <- function(yE, nE, yC, nC, w0 = 0.50) {
  aH <- 1 + yE
  bH <- 1 + nE - yE
  log_LH <- lchoose(nC, yC) +
    lbeta(aH + yC, bH + nC - yC) - lbeta(aH, bH)
  log_LV <- lchoose(nC, yC) +
    lbeta(1 + yC, 1 + nC - yC) - lbeta(1, 1)
  wpost <- plogis(qlogis(w0) + log_LH - log_LV)
  list(wpost = wpost,
       hist = c(a = aH + yC, b = bH + nC - yC),
       vague = c(a = 1 + yC, b = 1 + nC - yC))
}

draw_rmp_control <- function(fit, K = 5000) {
  use_hist <- rbinom(K, 1, fit$wpost) == 1
  out <- numeric(K)
  out[use_hist] <- rbeta(sum(use_hist), fit$hist["a"], fit$hist["b"])
  out[!use_hist] <- rbeta(sum(!use_hist), fit$vague["a"], fit$vague["b"])
  out
}

qtb_binary <- function(status,
  yT, nT, yC, nC, yE = NULL, nE = NULL,
  yCx = NULL, nCx = NULL, yEx = NULL, nEx = NULL,
  target_w = c(0.5, 0.5), w0 = 0.50, K = 5000,
  cstar = NULL) {

  status <- match.arg(status,
    c("qualified", "repairable", "not_qualified"))
  pT <- rbeta(K, 1 + yT, 1 + nT - yT)

  if (status == "not_qualified") {
    pC <- rbeta(K, 1 + yC, 1 + nC - yC)
    hist_wt_summary <- 0
  }

  if (status == "qualified") {
    fit <- rmp_beta_binom(yE, nE, yC, nC, w0)
    pC <- draw_rmp_control(fit, K)
```

```
    hist_wt_summary <- fit$wpost
  }

  if (status == "repairable") {
    stopifnot(length(yCx) == 2, length(nCx) == 2,
              length(yEx) == 2, length(nEx) == 2,
              length(target_w) == 2,
              abs(sum(target_w) - 1) < 1e-12)
    pCx <- vector("list", 2)
    wpost_x <- numeric(2)
    for (j in 1:2) {
      if (nEx[j] == 0) {
        pCx[[j]] <- rbeta(K, 1 + yCx[j], 1 + nCx[j] - yCx[j])
        wpost_x[j] <- 0
      } else {
        fit <- rmp_beta_binom(yEx[j], nEx[j], yCx[j], nCx[j], w0)
        pCx[[j]] <- draw_rmp_control(fit, K)
        wpost_x[j] <- fit$wpost
      }
    }
    pC <- target_w[1] * pCx[[1]] + target_w[2] * pCx[[2]]
    hist_wt_summary <- sum(target_w * wpost_x)
  }

  delta <- pT - pC
  ans <- list(
    posterior_mean = mean(delta),
    credible_interval = unname(quantile(delta, c(0.025, 0.975))),
    Pr_delta_gt_0 = mean(delta > 0),
    historical_component_weight_summary = hist_wt_summary)
  if (!is.null(cstar)) ans$success <- ans$Pr_delta_gt_0 > cstar
  ans
}

# Qualified source (fill in observed counts):
# qtb_binary("qualified", yT, nT, yC, nC, yE, nE,
#            cstar = 0.97236254488695761)

# Not-qualified source: external data are intentionally omitted.
# qtb_binary("not_qualified", yT, nT, yC, nC,
#            cstar = 0.97236254488695761)

# Repairable binary-X source: supply counts for X=0 and X=1.
# qtb_binary("repairable", yT, nT, yC, nC,
#            yCx = c(yC0, yC1), nCx = c(nC0, nC1),
#            yEx = c(yE0, yE1), nEx = c(nE0, nE1),
#            target_w = c(0.5, 0.5),
#            cstar = 0.97236254488695761)
```

**Interpretation.** The same source counts can lead to different primary analyses because Steps 1–4 determine what external information reaches Step 5 and in what form. For a qualified source, the code applies the RMP directly to the original external counts. For a repairable source, the same RMP is applied through the specified conditional analysis before target-population standardization. For a source not qualified for primary borrowing, the external counts never enter the primary posterior. Step 5 then determines borrowing strength for the information that remains.

## Supplementary Illustration 1. Applying QTB to a continuous endpoint

**Purpose.** This worked illustration shows how the five-step QTB framework can be implemented for a simple continuous endpoint. It is included to demonstrate portability beyond the binary Beta-Binomial example. It was not part of the simulation study, and no operating-characteristic claim is made for this setting.

Illustrative setup. Let Y be a continuous outcome for which larger values are favorable. Let mu_T and mu_C denote the marginal treatment and control means in the randomized-trial target population. To keep the calculation transparent, assume a known common standard deviation sigma. Let ybar_E, ybar_C, and ybar_T denote the external-control, concurrent-control, and treatment-arm sample means, with sample sizes n_E, n_C, and n_T. The same routing logic would apply with unknown variance using a normal-inverse-gamma model or another prespecified Bayesian analysis.

$$\Delta_R = \mu_T - \mu_C, \qquad V_E = \frac{\sigma^2}{n_E}, \qquad V_C = \frac{\sigma^2}{n_C}$$

The three Step-4 dispositions remain unchanged.

- **Qualified:** the external and current data use the same outcome instrument, assessment time, and target definition, and no identified material difference requires repair or exclusion. The external information proceeds unchanged to Step 5.
- **Repairable:** a measured prognostic factor X has a different distribution in the external and target populations, but X is observed in both sources with adequate overlap. The repair specifies borrowing within X strata followed by standardization of the stratum-specific posterior control means to the target-population distribution.
- **Not qualified for primary borrowing:** the external source measures a different continuous construct or scale and no validated crosswalk or patient-level information is available to reconstruct the target outcome. No external information proceeds to Step 5 for the primary analysis.

Step 5 with a Normal-Normal robust mixture prior.

For a qualified source, use the external mean as the center of an informative historical component. With V_E = sigma^2/n_E and V_C = sigma^2/n_C, let

Historical component:

$$\mu_C \sim N(m_E, V_E), \qquad m_E = \bar{Y}_E$$

Vague component:

$$\mu_C \sim N(m_0, V_0)$$

where m_0 and V_0 are prespecified to give a weakly informative or vague component. The current-control sampling model is

$$\bar{Y}_C \mid \mu_C \sim N(\mu_C, V_C)$$

The predictive support for the current data under the two components is

$$L_H = f_N(\bar{Y}_C;\, m_E,\, V_E + V_C)$$

$$L_V = f_N(\bar{Y}_C;\, m_0,\, V_0 + V_C)$$

The posterior historical-component weight is

$$w_{\text{post}} = \frac{w_0 L_H}{w_0 L_H + (1 - w_0) L_V}$$

Thus, as in the binary RMP, Step 5 does not re-decide the classification from Steps 1–4. The selected continuous-outcome borrowing model determines how strongly the external information passed forward from Step 4 contributes according to the residual empirical compatibility or source uncertainty that remains.

The two posterior component distributions are Normal. For the historical component,

$$V_{H,\text{post}} = (V_E^{-1} + V_C^{-1})^{-1}, \qquad m_{H,\text{post}} = V_{H,\text{post}} \left( \frac{m_E}{V_E} + \frac{\bar{Y}_C}{V_C} \right)$$

For the vague component,

$$V_{V,\text{post}} = (V_0^{-1} + V_C^{-1})^{-1}, \qquad m_{V,\text{post}} = V_{V,\text{post}} \left( \frac{m_0}{V_0} + \frac{\bar{Y}_C}{V_C} \right)$$

Posterior draws for mu_C are then sampled from the two-component mixture using w_post. The treatment mean is estimated from the randomized treatment arm only, and draws of Delta_R = mu_T - mu_C provide the posterior treatment-effect distribution.

Repairable population shift.

If X is a binary prognostic factor and the source is classified as repairable, apply the same Normal-Normal RMP separately for X=0 and X=1. Let mu_C0^(b) and mu_C1^(b) be posterior draws from the two stratum-specific analyses and let pi_R0 and pi_R1 be the prespecified target-population proportions. Standardize by

$$\mu_C^{R,(b)} = \pi_{R0} \mu_{C0}^{(b)} + \pi_{R1} \mu_{C1}^{(b)}$$

and then form

$$\Delta_R^{(b)} = \mu_T^{(b)} - \mu_C^{R,(b)}$$

This is the continuous-outcome analogue of the World-2 repair: the known population-composition difference determines the within-stratum comparison and target-population standardization structure before residual empirical compatibility is used to determine borrowing strength.

Minimal R illustration.

The following base-R code implements the qualified continuous-outcome route under a known common sigma. It is intentionally short and is not production software. The QTB status must already have been assigned before this function is called.

```
normal_rmp <- function(ybarE, nE, ybarC, nC, sigma,
                       w0 = 0.50, m0 = 0, V0 = 100^2) {
  VE <- sigma^2 / nE
  VC <- sigma^2 / nC

  log_LH <- dnorm(ybarC, mean = ybarE,
                  sd = sqrt(VE + VC), log = TRUE)
  log_LV <- dnorm(ybarC, mean = m0,
```

```
                  sd = sqrt(V0 + VC), log = TRUE)
  wpost <- plogis(qlogis(w0) + log_LH - log_LV)

  VH <- 1 / (1 / VE + 1 / VC)
  mH <- VH * (ybarE / VE + ybarC / VC)
  VV <- 1 / (1 / V0 + 1 / VC)
  mV <- VV * (m0 / V0 + ybarC / VC)

  list(wpost = wpost,
       hist = c(mean = mH, var = VH),
       vague = c(mean = mV, var = VV))
}

# Example of the qualified route:
# fit <- normal_rmp(ybarE = 0.00, nE = 100,
#                   ybarC = 0.10, nC = 25, sigma = 1)
# fit$wpost

# For a repairable binary-X source, run normal_rmp()
# separately within X=0 and X=1, draw each posterior,
# then standardize the draws using the target weights.
# For a not-qualified source, do not call normal_rmp();
# use the prespecified trial-only primary analysis.
```

**Interpretation.** This illustration changes the likelihood and conjugate updating from Beta-Binomial to Normal-Normal, but it does not change the QTB architecture. Steps 1–4 determine what external information reaches Step 5 and in what form; Step 5 then determines borrowing strength. Because this example was not evaluated in the simulation study, it should be read as an implementation illustration rather than evidence about operating characteristics.

## Supplementary Illustration 2. Applying QTB to a time-to-event endpoint

**Purpose.** This worked illustration shows how the QTB workflow can be implemented in a deliberately simple time-to-event setting. It is included to make the framework portable beyond binary and continuous outcomes. It was not part of the simulation study, and no operating-characteristic claim is made for this setting. The exponential model is used only for transparency; it is not proposed as a default survival model for clinical trials.

Illustrative target. Let T denote event time and suppose smaller event hazards are favorable. For treatment arm a in {T,C}, assume an exponential model with hazard lambda_a. At a prespecified horizon tau, use the target-population survival difference as the illustrative estimand. The treatment arm remains randomized-trial only; candidate external information can inform only the control distribution.

$$S_a(t) = \exp(-\lambda_a t), \qquad \Delta_S(\tau) = S_T(\tau) - S_C(\tau)$$

The same three Step-4 dispositions apply before the time-to-event borrowing model is used.

**Qualified:** the external and current controls use the same time origin, event definition, and analysis strategy for relevant intercurrent events; follow-up and censoring are sufficiently aligned for the target analysis; and no identified material difference requires repair or exclusion. The external information proceeds unchanged to Step 5.

**Repairable:** a measured prognostic factor X has a different distribution in the external and target populations, but X is observed in both sources with adequate overlap. The repair specifies borrowing within X strata followed by standardization of stratum-specific survival to the target-population distribution.

**Not qualified for primary borrowing:** the target analysis starts follow-up at enrollment or randomization, but the external source contains only patients who reached treatment initiation and omitted pre-treatment failures cannot be recovered; or the event definition required by the target estimand cannot be reconstructed. Similar observed survival summaries do not resolve this mismatch, so no external information proceeds to Step 5 for the primary analysis.

Step 5 with an exponential-Gamma robust mixture prior.

For external information that proceeds to Step 5, let d_E and T_E denote the number of observed external-control events and total external-control person-time, and let d_C and T_C denote the corresponding concurrent-control quantities. Under an exponential model with noninformative censoring, use the Gamma(shape, rate) parameterization and the control likelihood

$$L(\lambda_C \mid d, T) \propto \lambda_C^d \exp(-\lambda_C T)$$

Start with a prespecified baseline Gamma(a_0, b_0) prior for the historical component. After the external data, the informative component is

Historical component:

$$\lambda_C \sim \text{Gamma}(a_H, b_H), \qquad a_H = a_0 + d_E, \qquad b_H = b_0 + T_E$$

Use a prespecified weak component Gamma(a_V, b_V) as the robust alternative. Conditional on the current exposure T_C, the integrated predictive support for d_C events under a Gamma(a,b) component is proportional to

$$L(a, b; d_C, T_C) \propto \frac{b^a \Gamma(a + d_C)}{\Gamma(a)(b + T_C)^{a + d_C}}$$

Therefore define L_H = L(a_H,b_H; d_C,T_C) and L_V = L(a_V,b_V; d_C,T_C). Common likelihood factors cancel in the ratio. The posterior historical-component weight is

$$w_{\text{post}} = \frac{w_0 L_H}{w_0 L_H + (1 - w_0) L_V}$$

The resulting control posterior is the two-component Gamma mixture

$$\lambda_C \mid D \sim w_{\text{post}} \text{Gamma}(a_H + d_C, b_H + T_C) + \left(1 - w_{\text{post}}\right) \text{Gamma}(a_V + d_C, b_V + T_C)$$

Posterior draws of lambda_C are sampled from this mixture. A trial-only posterior for lambda_T supplies treatment-arm draws, which are transformed as follows:

$$S_C(\tau) = e^{-\lambda_C \tau}, \qquad S_T(\tau) = e^{-\lambda_T \tau}, \qquad \Delta_S(\tau) = S_T(\tau) - S_C(\tau)$$

As in the binary and continuous examples, Step 5 determines borrowing strength for the external information passed forward from Steps 1–4; it does not re-decide the pre-borrowing classification.

Repairable population shift.

If X is a binary prognostic factor and the source is classified as repairable, apply the exponential-Gamma RMP separately within X=0 and X=1. On posterior draw b, let lambda_C0^(b) and lambda_C1^(b) denote the stratum-specific control hazards and let pi_R0 and pi_R1 be the prespecified target-population proportions. Standardize survival, not the hazards:

$$S_C^{R,(b)}(\tau) = \pi_{R0} e^{-\lambda_{C0}^{(b)} \tau} + \pi_{R1} e^{-\lambda_{C1}^{(b)} \tau}$$

then form

$$\Delta_S^{(b)}(\tau) = S_T^{(b)}(\tau) - S_C^{R,(b)}(\tau)$$

This is the time-to-event analogue of the World-2 repair: the known population-composition difference determines the within-stratum borrowing and target-population standardization structure before residual empirical compatibility is used to determine borrowing strength.

Minimal R illustration.

The following base-R code implements the qualified exponential route. It is intentionally short and is not production software. Prior hyperparameters must be prespecified on the chosen time scale, and the QTB status must already have been assigned before this function is called.

```
tte_rmp <- function(dE, TE, dC, TC,
                    a_base, b_base, a_vague, b_vague,
                    w0 = 0.50) {
  stopifnot(TE > 0, TC > 0, dE >= 0, dC >= 0,
            w0 > 0, w0 < 1)

  aH <- a_base + dE
  bH <- b_base + TE
```

```
  log_ml <- function(a, b, d, TT) {
    a * log(b) - lgamma(a) + lgamma(a + d) -
      (a + d) * log(b + TT)
  }

  log_LH <- log_ml(aH, bH, dC, TC)
  log_LV <- log_ml(a_vague, b_vague, dC, TC)
  wpost <- plogis(qlogis(w0) + log_LH - log_LV)

  list(
    wpost = wpost,
    hist  = c(shape = aH + dC, rate = bH + TC),
    vague = c(shape = a_vague + dC, rate = b_vague + TC)
  )
}

draw_control_survival <- function(fit, tau, B = 5000) {
  z <- rbinom(B, 1, fit$wpost)
  lambda <- numeric(B)
  lambda[z == 1] <- rgamma(sum(z == 1),
                           shape = fit$hist['shape'],
                           rate  = fit$hist['rate'])
  lambda[z == 0] <- rgamma(sum(z == 0),
                           shape = fit$vague['shape'],
                           rate  = fit$vague['rate'])
  exp(-lambda * tau)
}

# Qualified route:
# fitC <- tte_rmp(dE, TE, dC, TC,
#                 a_base, b_base, a_vague, b_vague)
# SC <- draw_control_survival(fitC, tau = tau)

# Repairable binary-X route: run tte_rmp() separately
# within X=0 and X=1, draw S_C0(tau) and S_C1(tau),
# then standardize survival using target weights.
# Not-qualified route: do not call tte_rmp(); use the
# prespecified trial-only primary analysis.
```

**Boundary conditions.** This example assumes exponential event times within the relevant analysis group, a prespecified fixed horizon tau, and noninformative censoring so that event counts and total person-time provide the simple likelihood above. It does not address nonproportional hazards, interval-censored progression, informative censoring, competing risks, recurrent events, complex intercurrent-event strategies, or reconstruction of incompatible time origins. Those settings require an estimand and survival model appropriate to the scientific question. The purpose here is only to show that the QTB routing logic can precede a standard time-to-event borrowing model; it is not evidence that this particular exponential implementation is adequate for general survival endpoints.

**Interpretation.** The likelihood and conjugate update have changed from Beta-Binomial or Normal-Normal to exponential-Gamma, but the QTB architecture has not. Steps 1–4 determine what time-to-event information reaches Step 5 and in what form; Step 5 then determines borrowing strength. Because this illustration was not evaluated in the simulation study, it should be read as an implementation example rather than validation of time-to-event operating characteristics.

## S6. Posterior computation and Monte Carlo stability

The World-2 repair propagates uncertainty from the two stratum-specific posterior distributions through the 50%/50% target-population standardization step using posterior Monte Carlo draws. We compared K=1000, 2000, and 5000 draws and retained the smallest candidate that satisfied all prespecified stability criteria relative to the K=5000 reference. The criteria were: 95th percentile absolute posterior-probability difference ≤0.005; 95th percentile absolute posterior-mean treatment-effect difference ≤0.002; and 95th percentile absolute difference for each 95% credible-interval endpoint ≤0.01. Neither K=1000 nor K=2000 satisfied all criteria, so K=5000 was used in the final analysis. This choice was based on numerical stability, not on scientific operating-characteristic performance.

**Supplementary Table S1. Posterior Monte Carlo stability assessment**

| K | p95 \|Δ posterior probability\| | p95 \|Δ posterior mean\| | p95 \|Δ lower CI\| | p95 \|Δ upper CI\| | Result |
|---|---|---|---|---|---|
| 1000 | 0.022620 | 0.0055128 | 0.0157653 | 0.0135872 | Not satisfied |
| 2000 | 0.012835 | 0.0034533 | 0.0112078 | 0.0088313 | Not satisfied |
| 5000 | 0 | 0 | 0 | 0 | Satisfied (reference) |

**Note:** The K=5000 row is the reference comparison and therefore has zero differences by construction; it should not be interpreted as evidence of zero posterior Monte Carlo error.

## S7. Exact RCT-only calibration of the success threshold

Under the prespecified null, Y_T~Binomial(50,0.30) and Y_C~Binomial(25,0.30). All 51×26=1,326 possible randomized treatment-control response-count pairs were enumerated. For each pair, Q(y_T,y_C)=P(p_T>p_C | y_T,y_C) was calculated under independent Beta(1,1) priors. Success is declared when Q>c*. The exact null rejection probability for a candidate cutoff is the joint-binomial probability mass of all count pairs satisfying the decision rule.

The prespecified cutoff is c*=0.97236254488695761, yielding exact RCT-only Type I error 0.022539846668296316. The immediately less stringent attainable cutoff, 0.97012764644839533, yields Type I error 0.025730913848454183 and therefore exceeds the one-sided 0.025 target. Thus c* is the least stringent discrete cutoff satisfying the prespecified pointwise requirement under the strict > rule. The same full-precision c* is used for M0–M3; the main text reports the rounded value c*=0.972. This calibration applies to p_T=p_C=0.30 and does not imply uniform Type I-error control over every possible null configuration.

## S8. Simulation scenario families

World 1 (fully compatible evidence): P_E(X=1)=0.50, and within each X stratum the external and randomized-trial controls have the same response probability. No identified material difference requires repair or exclusion, so the source is classified as qualified and the external information proceeds unchanged to Step 5.

World 2 (repairable population shift): P_E(X=1)=0.80 while the target population has P_R(X=1)=0.50. Within each X stratum, the external and current controls have the same response probability by construction; this equality is part of the data-generating mechanism and is not used to determine classification. Because X is measured in both datasets and both strata are represented, the source is classified as repairable and Step 5 uses the stratified RMP and target-population standardization described in S5.

World 3 (qualified external data with residual outcome drift): $P_E(X=1)=0.50$, and Steps 1–4 identify no material population, endpoint, measurement, or available-data difference requiring repair or exclusion. For data generation only, an external logit offset produces marginal external response probabilities of 0.15, 0.20, 0.25, 0.30, 0.35, 0.40, and 0.45. This drift is not treated as information available to Steps 1–4. The source is therefore classified as qualified, the external information proceeds unchanged to Step 5, and the RMP responds to the residual outcome disagreement.

World 4 (known endpoint mismatch): the true external response follows the same control model as the randomized trial, but some true nonresponses are recorded as responses: $P(Y_E^*=1 \mid Y_E=1)=1$ and $P(Y_E^*=1 \mid Y_E=0)=q$, with $q\in\{0.05,0.10,0.20\}$. The presence of a material endpoint-recording problem is treated as known, but the information needed to reconstruct the target endpoint is unavailable. The source is therefore classified as not qualified for primary borrowing, and no external information proceeds to Step 5. The value of q is used only to generate the simulated outcomes and is not provided as correction information to the analysis.

World 5 (same outcome distribution, different QTB decisions): two external-control settings have the same recorded response probability, 0.37. In the residual-drift setting, Steps 1–4 identify no material difference requiring repair or exclusion, so the source is classified as qualified. In the endpoint-mismatch setting, the true response probability is 0.30 and $q=0.10$ misclassification also produces a recorded response probability of 0.37; the known, nonrepairable endpoint problem leads to a not-qualified classification. Thus, the observed number of external responses has the same $\text{Binomial}(n_E,0.37)$ distribution in both settings, while the prespecified scientific information leads to different pre-borrowing actions.

Worlds 1 and 2 use $n_E=25,50,100,250$ under both null and alternative treatment conditions. Worlds 3 and 4 use $n_E=50,100,250$ under both conditions. World 5 uses $n_E=50,100,250$ under the null only. Crossing these settings with the prespecified scenario parameters yields 82 scenarios in the final simulation design.

## S9. Analysis strategies

M0 uses randomized-trial data only. M1 fully pools the external and concurrent controls. M2 applies the prespecified RMP directly to the original external controls without QTB Steps 1–4. M3 uses the same RMP in Step 5 after routing through Steps 1–4: M3=M2 when the source is qualified and proceeds unchanged; M3 uses the World-2 repair analysis when the source is repairable; and M3=M0 when the source is not qualified for primary borrowing and no external information proceeds to Step 5. These equalities are definitions of the analysis methods, not simulation findings. External data never inform the treatment arm.

## S10. Operating-characteristic definitions

$$\text{Bias}_j = E(\hat{\Delta}_j) - \Delta_R$$

$$\text{RMSE}_j = \sqrt{E\left[\left(\hat{\Delta}_j - \Delta_R\right)^2\right]}$$

Coverage is the proportion of 95% posterior credible intervals containing $\Delta_R$. Mean interval width is the Monte Carlo mean of the upper minus lower credible-interval endpoints. Type I error is the probability of satisfying $P(\Delta_R>0 \mid D)>c^*$ under $\Delta_R=0$; power is the corresponding probability under $\Delta_R=0.20$. The same $c^*$ is used for all methods. For standard RMP analyses, the reported historical-component weight is the posterior probability assigned to the historical component. For the repaired World-2 analysis, the reported weight is the 50%/50% target-population average of the two stratum-specific posterior historical-component weights; a stratum with no external controls receives weight zero. These quantities summarize the RMP update and should not be interpreted as the literal fraction of external participants borrowed.

$$ER_j = \frac{\text{MSE}(M0)}{\text{MSE}(M_j)}$$

ER_j>1 indicates lower MSE than the trial-only analysis M0. Bias, RMSE, coverage, interval width, Type I error or power, historical-component weight, and ER are interpreted together.

## S11. Random-number generation and reproducibility

Deterministic random-number streams were assigned by scenario, replication, and computation step so that the data generated for any simulation unit did not depend on scenario order, method order, or serial versus parallel execution. Reproducibility was checked by repeating runs, reversing scenario order, reversing method order, and comparing serial with parallel execution. Parallelization changed execution only; it did not change the statistical procedure or the mapping of random-number streams.

## S12. Automated testing and quality control

Automated checks verified scenario completeness, replication counts, uniqueness of scenario-method-replication keys, absence of unexpected missing or non-finite values, agreement between posterior success probabilities and success indicators, uniqueness of random-number streams, exact routing identities, numerical calibration, posterior-computation checks, and reproducibility across execution modes. Qualified scenarios required M2=M3 exactly; scenarios not qualified for primary borrowing required M0=M3 exactly. Scientific operating characteristics were released only after all final checks passed.

## S13. Development versus final simulation

Development runs were used only for implementation checks, numerical-stability assessment, debugging, and software quality control. Development estimates of Type I error, power, bias, RMSE, coverage, interval width, historical-component weight, and efficiency were not treated as scientific results and were not used to choose favorable scenarios or alter the prespecified methods. All scientific results are based on the final R=10,000 simulation after the design and code had been frozen and all final quality-control checks had passed.

## S14. Reproducibility materials

The reproducibility archive is planned to include the prespecified scientific and simulation specification, final source code, deterministic random-number mapping, exact calibration code, posterior and RMP routines, repair and target-population standardization code, automated tests, final quality-control reports, raw final simulation outputs, and the summarized data used for manuscript tables and figures.

# Supplementary Appendix A. Clinically Grounded Worked QTB Examples in Pediatric Oncology

**Purpose.** These worked examples illustrate how QTB can be applied to clinically realistic pediatric-oncology problems that are more varied than the single population-shift repair used in the main simulation study. They are hypothetical, clinically grounded illustrations. The numbers are pedagogical and are not results from a real NANT, INRG, or other patient-level dataset. No operating-characteristic claim is made for these examples. Their purpose is to make the sequence concrete: identified material difference -> repair or exclusion -> target-aligned external information -> residual empirical compatibility -> Step-5 borrowing -> residual-bias assessment.

## A0. Repairability taxonomy for common external-data differences

**Supplementary Table A1. Repairability taxonomy for common differences between external and current data.**

| Problem | Repairable when... | Possible repair | Not adequately repairable when... |
|---|---|---|---|
| Population mix / prognostic imbalance | Key prognostic factors are measured in both sources and target support/overlap is adequate. | Stratification, weighting, standardization, or target-population outcome modeling. | Key factors are missing, overlap is poor, or repair requires unsupported extrapolation. |
| Eligibility mismatch | Variables needed to reapply the target eligibility criteria are retained. | Reapply target eligibility criteria to construct a target-aligned external cohort. | A key eligibility variable was not recorded or eligible and ineligible patients cannot be distinguished. |
| Time-zero mismatch | Earlier enrollment/leukapheresis/treatment-course dates and relevant events are retained. | Reconstruct target time zero, risk set, and follow-up before Step 5. | Only treated/infused patients remain and pre-treatment failures or omitted patients cannot be recovered. |
| Endpoint-definition mismatch | Component-level measurements, events, or dates required by the target definition are available. | Re-adjudicate or reconstruct the target endpoint. | Only an aggregate endpoint flag remains and the target endpoint cannot be reconstructed. |
| Response-confirmation mismatch | Serial scans, response categories, dates, and confirmation information are retained. | Recalculate confirmed ORR using the target confirmation rule. | Only best response is retained without follow-up confirmation information. |
| PFS assessment-schedule mismatch | Assessment dates and measurements support a prespecified target-window reconstruction or scientifically justified observation-process model. | Reconstruct the target-window endpoint or use a prespecified interval-censoring/observation-process analysis when appropriate. | Target-window disease status is unobserved and must be guessed or imputed without adequate identifying information. |
| Disease-component composition | Disease-site/component information relevant to the target response definition is measured with adequate support. | Stratify or reconstruct a common target response endpoint using component-level data. | Relevant disease components are missing or target support is absent. |
| Calendar-era / site mix | Measured drivers of era/site differences are available with adequate overlap. | Standardization, weighting, or modeling to the target distribution. | Important practice drift, supportive care, diagnostic intensity, or other source effects are largely unmeasured. |

Interpretation. Repairability is a property of the target, the source, and the information available for repair. The same apparent mismatch can therefore lead to different QTB classifications in different data sources.

## A1. Neuroblastoma endpoint reconstruction: best response through Cycle 4 versus target Cycle-2 ORR

**Target and known difference.** Consider a hypothetical relapsed/refractory neuroblastoma trial whose primary endpoint is Cycle-2 ORR, with a small concurrent control group and a historical NANT-like source that originally reports best response through Cycle 4. Illustratively, the historical source reports 50/100 responders (50%) using the broader Cycle-4 window, whereas current controls show 9/25 responders (36%) at Cycle 2. These quantities do not represent the same endpoint window, so the raw 50% versus 36% comparison should not be handed directly to Step 5.

**Why this can be repairable.** Assume the historical source retains the Cycle-2 assessment date, component-level disease measurements, and the information needed to apply the locked target response rule. Historical patients can then be re-adjudicated under the target Cycle-2 definition. If the source retained only a final best-response-through-Cycle-4 indicator, without the underlying Cycle-2 assessments, the same apparent mismatch might instead be not adequately repairable for this target.

**Step 4 repair.** Suppose target-aligned re-adjudication changes the historical response count from 50/100 best response through Cycle 4 to 31/100 Cycle-2 responders. The object passed to Step 5 is now the reconstructed 31/100 target-aligned endpoint, not the original 50/100 endpoint. This is endpoint reconstruction, not outcome matching: the repair rule is determined by the target definition and the information retained in the historical source.

**Residual empirical compatibility and Step 5.** After repair, the relevant empirical comparison is 31% versus 36%, not 50% versus 36%. The known endpoint-window difference has been addressed. The remaining 31% versus 36% disagreement may reflect sampling variation, temporal drift, site effects, unmeasured disease features, or other source effects. Using the same Beta-Binomial RMP as in the paper, these toy counts give a posterior historical-component weight of approximately 0.775. The important point is not the exact weight; it is that the Bayesian engine is now responding to target-aligned information.

**What happens if repair is skipped.** If the same RMP is applied directly to the wrong historical endpoint, 50/100 versus current 9/25, it can detect disagreement and downweight the historical component (approximately 0.633 for these toy counts), but it is still borrowing from the wrong endpoint information. Repair can therefore lead to more, not less, borrowing when the target-aligned information is more compatible. Repair is alignment; borrowing strength is determined afterward.

## A2. NANT-like relapsed/refractory neuroblastoma population shift

**Clinical setup.** Consider a hypothetical trial in which the current target population is 40% primary refractory and 60% relapsed, whereas a historical NANT-like control source is 70% refractory and 30% relapsed. Refractory/relapsed status is measured in both sources, is clinically prognostic for response, and has adequate support in both strata. The known population-composition difference is therefore material but potentially repairable.

**Illustrative data.** Historical controls contain 17/84 refractory responders (20.2%) and 16/36 relapsed responders (44.4%), for a raw marginal ORR of 33/120 = 27.5%. Current controls contain 2/10 refractory responders (20.0%) and 7/15 relapsed responders (46.7%), for a marginal ORR of 9/25 = 36.0%. A direct marginal comparison mixes the known difference in refractory/relapsed composition with whatever source disagreement remains.

**Step 4 repair.** Borrow within refractory and relapsed strata rather than marginally, then standardize to the target trial's 40%/60% composition. Using the historical stratum-specific rates alone, the target-aligned historical control response is 0.40(0.202) + 0.60(0.444) = approximately 34.8%, rather than the raw marginal 27.5%. A useful representation of the repaired external information is the pair of stratum-specific historical counts together with the target-population weights.

**Residual compatibility and Step 5.** After repair, Step 5 compares 20.2% versus 20.0% within the refractory stratum and 44.4% versus 46.7% within the relapsed stratum. Using the same RMP in each stratum, the toy posterior historical-component weights are approximately 0.758 and 0.731, respectively. Posterior draws are then standardized back to the target 40%/60% population. The same Bayesian engine is used; what changes is the information allowed to reach it and the form in which that information is compared.

**Interpretation.** The repair does not claim that historical and current patients are exchangeable. It removes one identified population-mix difference from the comparison structure. Remaining within-stratum disagreement is what Step 5 sees as residual empirical compatibility, and residual bias may still remain after borrowing.

## A3. Time-zero mismatch: the same apparent problem can be repairable or not qualified

**Repairable version.** Suppose the target analysis starts follow-up at enrollment or leukapheresis, but an external cellular-therapy registry is usually summarized from infusion. If the registry still contains enrollment/leukapheresis dates, bridging-therapy information, manufacturing failures, pre-infusion progression, and never-infused patients, the target risk set and target time zero can potentially be reconstructed before Step 5.

**Not-qualified version.** If the external source contains only infused patients and the omitted pre-infusion failures or earlier treatment course cannot be recovered, the target estimand cannot be reconstructed. The source would then be not qualified for primary borrowing for that target. Similar observed survival after infusion does not repair the missing risk set.

**Key point.** Repairability is a property of the target, the source, and the information actually retained. A label such as “time-zero mismatch” does not by itself determine the QTB classification.

## A4. Response-confirmation mismatch

**Known difference and repair.** Suppose the target endpoint is confirmed ORR, requiring CR/PR confirmation at least 4 weeks later, while an external source originally reports unconfirmed best response. If serial scan dates, tumor measurements, response categories, and confirmation information are retained, the historical endpoint can be reconstructed under the target rule. Illustratively, an external 40/80 = 50% unconfirmed ORR might become 28/80 = 35% confirmed ORR after reconstruction.

**Residual compatibility.** If current controls show 7/25 = 28% confirmed ORR, the relevant post-repair comparison is 35% versus 28%. That remaining difference is not evidence that the repair failed; the endpoint definition has been aligned, and Step 5 now responds to the residual empirical agreement or disagreement.

## A5. PFS assessment schedule: a boundary case for repairability

**Potentially repairable patient.** Suppose the target trial images every 8 weeks +/- 2 weeks. An external patient has no progression at week 7.5 and progression at week 16.5. If scan dates, component-level measurements, and the common progression rule are available, the observations can be mapped to the target week-8 and week-16 windows and re-adjudicated under the target rule.

**Potentially nonrepairable patient.** Another external patient has no progression at week 12 and progression at week 24, while the target requires a week-16 assessment. The week-16 disease status is unobserved. It should not be manufactured by treating “no scan” as “no progression” or by arbitrary interpolation. A prespecified interval-censoring or observation-process model may sometimes provide a scientifically justified route, but if the target endpoint depends on an unrecoverable fixed-window status, the difference may not be adequately repairable.

**Key point.** Different assessment schedules are not automatically repairable merely because assessment dates are available. QTB requires the repair itself to be scientifically justified and supported by information in the source.

### A6. Eligibility mismatch

Target-aligned cohort construction. Suppose the current trial requires measurable disease, no more than two prior lines of therapy, and a prespecified age/risk profile, while the external registry includes a broader population. If the variables needed to reapply those criteria are recorded, the external cohort can be reconstructed to match the target eligibility definition before Step 5. If a critical eligibility variable is absent, the mismatch may not be adequately repairable. After eligibility alignment, historical and current outcomes can still differ, and that residual disagreement is what Step 5 evaluates.

### A7. Disease-component composition and response assessment

Target-compatible disease representation. Neuroblastoma response can depend on the pattern of measurable soft-tissue disease, bone disease, marrow disease, and other disease components. If the historical and current sources differ materially in this composition but the component-level disease information required by the target response rule is retained with adequate support, QTB can define a target-compatible stratification or endpoint reconstruction before borrowing. If relevant disease components are missing, a nominally similar response rate does not establish that the endpoint is comparable.

### A8. Calendar-era and site mix

Measured versus unmeasured era differences. A historical source may differ from the current trial in treatment backbone, staging, supportive care, imaging practice, or site mix. When the important drivers are measured and overlap is adequate, standardization or weighting may address the measured part of the difference. Unmeasured practice drift cannot be declared repaired merely because the measured variables have been aligned. It remains a residual source concern for Step 5 and sensitivity analysis.

### A9. Practical take-home

- Repair known differences before borrowing.
- Borrow on target-aligned external information rather than automatically on the raw external dataset.
- Residual empirical compatibility begins where identifiable repair ends; qualification or repair does not certify exchangeability.
- The same apparent mismatch can be repairable in one source and not qualified in another, depending on whether the information required for repair is actually available.
- The Step-5 borrowing engine may be unchanged even when QTB changes the resulting analysis, because Steps 1-4 change what reaches that engine and in what form.

## Supplementary Appendix B. Practical QTB Documentation Template for Real Applications

Purpose. This template is intended to make the path from scientific assessment to borrowing explicit and auditable in a real application. It does not replace disease-specific scientific judgment or a prespecified statistical analysis plan.

**Supplementary Table B1. Suggested documentation fields for a QTB application.**

| Field | What to document |
|---|---|
| 1. Target | Target population, treatment/control condition, endpoint, time zero, intercurrent-event strategy, and population-level estimand. |
| 2. Candidate external source | Source, calendar period, sites, data provenance, and intended role in the target analysis. |
| 3. Identified material difference | The concrete difference relative to the target; avoid vague labels when the actionable feature can be named. |
| 4. Why the difference matters | How the difference could alter the target comparison, risk set, endpoint, or interpretation. |
| 5. Information available for repair | Variables, dates, component-level measurements, overlap/support, or validation information actually retained in the source. |
| 6. QTB classification | Qualified, repairable, or not qualified for primary borrowing, with the rationale documented before observed outcome agreement determines borrowing strength. |
| 7. Step-4 action | Use directly, specify the exact repair, or exclude from primary borrowing. |
| 8. External information passed to Step 5 | Write down D_E* explicitly: original data, target-aligned/reconstructed representation, or no external information. |
| 9. Residual empirical compatibility | Describe what empirical agreement/disagreement remains after identified material differences have been addressed. |
| 10. Step-5 borrowing engine | Specify the Bayesian borrowing model, tuning parameters, and reporting quantity used to determine borrowing strength. |
| 11. Residual-bias / sensitivity plan | State what source differences could remain unrecognized or incompletely repaired and the prespecified sensitivity or bias-aware analyses used to assess them. |

Reporting guardrails. Repair should be defined from the target and the information available in the source, not chosen to make outcomes look more similar. Residual empirical disagreement is not the same as residual bias. Qualification or repair does not certify exchangeability. Step 5 does not re-decide the Step-3 classification, and the posterior historical-component weight is not a literal fraction of external participants borrowed.

A concise way to summarize the workflow is: repair known differences before borrowing; borrow on target-aligned information rather than automatically on the raw external dataset; and treat residual empirical compatibility as beginning where identifiable repair ends.

## Supplementary Results

All results below come from the final R=10,000 simulation. Development runs are not included. Posterior Monte Carlo stability and exact threshold calibration are reported in the Supplementary Methods because they are numerical-validation and design quantities rather than scientific simulation results. Across Supplementary Tables S2–S6, pE(obs) denotes the recorded marginal external response probability. Hist. wt. denotes the posterior historical-component weight, or the prespecified target-population summary of stratum-specific weights for the repair analysis; it is not the literal fraction of external participants borrowed. ER=MSE(M0)/MSE(method), so values greater than 1 indicate lower MSE than the trial-only analysis.

### Supplementary Table S2. World 1 - Fully compatible evidence

| Truth | nE | pE(obs) | Method | Bias | RMSE | Coverage | Mean width | Type I error | Power | MCSE | Hist. wt. | ER |
|---|---|---|---|---|---|---|---|---|---|---|---|---|
| Null | 25 | 0.300 | M0 | -0.0065 | 0.1055 | 0.9539 | 0.4166 | 0.0215 | | 0.0015 | 0.0000 | 1.0000 |
| Null | 25 | 0.300 | M1 | 0.0001 | 0.0877 | 0.9509 | 0.3474 | 0.0274 | | 0.0016 | | 1.4486 |
| Null | 25 | 0.300 | M2 | -0.0022 | 0.0940 | 0.9617 | 0.3871 | 0.0199 | | 0.0014 | 0.6595 | 1.2604 |
| Null | 25 | 0.300 | M3 | -0.0022 | 0.0940 | 0.9617 | 0.3871 | 0.0199 | | 0.0014 | 0.6595 | 1.2604 |
| Alt | 25 | 0.300 | M0 | -0.0130 | 0.1099 | 0.9530 | 0.4286 | | 0.4123 | 0.0049 | 0.0000 | 1.0000 |
| Alt | 25 | 0.300 | M1 | -0.0061 | 0.0931 | 0.9489 | 0.3619 | | 0.5518 | 0.0050 | | 1.3937 |
| Alt | 25 | 0.300 | M2 | -0.0085 | 0.0991 | 0.9601 | 0.3992 | | 0.4661 | 0.0050 | 0.6631 | 1.2281 |
| Alt | 25 | 0.300 | M3 | -0.0085 | 0.0991 | 0.9601 | 0.3992 | | 0.4661 | 0.0050 | 0.6631 | 1.2281 |
| Null | 50 | 0.300 | M0 | -0.0078 | 0.1047 | 0.9536 | 0.4167 | 0.0201 | | 0.0014 | 0.0000 | 1.0000 |
| Null | 50 | 0.300 | M1 | 0.0020 | 0.0803 | 0.9535 | 0.3184 | 0.0260 | | 0.0016 | | 1.6990 |
| Null | 50 | 0.300 | M2 | -0.0015 | 0.0883 | 0.9652 | 0.3720 | 0.0174 | | 0.0013 | 0.6900 | 1.4075 |
| Null | 50 | 0.300 | M3 | -0.0015 | 0.0883 | 0.9652 | 0.3720 | 0.0174 | | 0.0013 | 0.6900 | 1.4075 |
| Alt | 50 | 0.300 | M0 | -0.0160 | 0.1099 | 0.9493 | 0.4285 | | 0.4003 | 0.0049 | 0.0000 | 1.0000 |
| Alt | 50 | 0.300 | M1 | -0.0063 | 0.0849 | 0.9511 | 0.3343 | | 0.6225 | 0.0048 | | 1.6744 |
| Alt | 50 | 0.300 | M2 | -0.0095 | 0.0934 | 0.9609 | 0.3860 | | 0.4994 | 0.0050 | 0.6866 | 1.3852 |
| Alt | 50 | 0.300 | M3 | -0.0095 | 0.0934 | 0.9609 | 0.3860 | | 0.4994 | 0.0050 | 0.6866 | 1.3852 |
| Null | 100 | 0.300 | M0 | -0.0082 | 0.1058 | 0.9509 | 0.4171 | 0.0240 | | 0.0015 | 0.0000 | 1.0000 |

| Truth | nE | pE(obs) | Method | Bias | RMSE | Coverage | Mean width | Type I error | Power | MCSE | Hist. wt. | ER |
|---|---|---|---|---|---|---|---|---|---|---|---|---|
| Null | 100 | 0.300 | M1 | 0.0036 | 0.0747 | 0.9459 | 0.2920 | 0.0307 | | 0.0017 | | 2.0068 |
| Null | 100 | 0.300 | M2 | -0.0006 | 0.0844 | 0.9636 | 0.3595 | 0.0196 | | 0.0014 | 0.7045 | 1.5709 |
| Null | 100 | 0.300 | M3 | -0.0006 | 0.0844 | 0.9636 | 0.3595 | 0.0196 | | 0.0014 | 0.7045 | 1.5709 |
| Alt | 100 | 0.300 | M0 | -0.0166 | 0.1104 | 0.9491 | 0.4288 | | 0.3970 | 0.0049 | 0.0000 | 1.0000 |
| Alt | 100 | 0.300 | M1 | -0.0039 | 0.0792 | 0.9519 | 0.3093 | | 0.7120 | 0.0045 | | 1.9424 |
| Alt | 100 | 0.300 | M2 | -0.0091 | 0.0897 | 0.9654 | 0.3742 | | 0.5379 | 0.0050 | 0.7010 | 1.5143 |
| Alt | 100 | 0.300 | M3 | -0.0091 | 0.0897 | 0.9654 | 0.3742 | | 0.5379 | 0.0050 | 0.7010 | 1.5143 |
| Null | 250 | 0.300 | M0 | -0.0085 | 0.1047 | 0.9563 | 0.4168 | 0.0191 | | 0.0014 | 0.0000 | 1.0000 |
| Null | 250 | 0.300 | M1 | 0.0060 | 0.0684 | 0.9499 | 0.2672 | 0.0308 | | 0.0017 | | 2.3466 |
| Null | 250 | 0.300 | M2 | 0.0002 | 0.0801 | 0.9688 | 0.3469 | 0.0168 | | 0.0013 | 0.7145 | 1.7114 |
| Null | 250 | 0.300 | M3 | 0.0002 | 0.0801 | 0.9688 | 0.3469 | 0.0168 | | 0.0013 | 0.7145 | 1.7114 |
| Alt | 250 | 0.300 | M0 | -0.0152 | 0.1100 | 0.9493 | 0.4288 | | 0.4055 | 0.0049 | 0.0000 | 1.0000 |
| Alt | 250 | 0.300 | M1 | -0.0010 | 0.0733 | 0.9517 | 0.2864 | | 0.7897 | 0.0041 | | 2.2521 |
| Alt | 250 | 0.300 | M2 | -0.0063 | 0.0852 | 0.9685 | 0.3621 | | 0.5873 | 0.0049 | 0.7144 | 1.6658 |
| Alt | 250 | 0.300 | M3 | -0.0063 | 0.0852 | 0.9685 | 0.3621 | | 0.5873 | 0.0049 | 0.7144 | 1.6658 |

*Note: M2 and M3 are identical by design because no identified material difference requires repair or exclusion; the external information proceeds unchanged to Step 5.*

## Supplementary Table S3. World 2 - Repairable population shift

| Truth | nE | pE(obs) | Method | Bias | RMSE | Coverage | Mean width | Type I error | Power | MCSE | Hist. wt. | ER |
|---|---|---|---|---|---|---|---|---|---|---|---|---|
| Null | 25 | 0.343 | M0 | -0.0069 | 0.1045 | 0.9562 | 0.4169 | 0.0222 | | 0.0015 | 0.0000 | 1.0000 |
| Null | 25 | 0.343 | M1 | -0.0196 | 0.0907 | 0.9459 | 0.3506 | 0.0157 | | 0.0012 | | 1.3286 |
| Null | 25 | 0.343 | M2 | -0.0115 | 0.0947 | 0.9607 | 0.3892 | 0.0161 | | 0.0013 | 0.6476 | 1.2186 |
| Null | 25 | 0.343 | M3 | -0.0130 | 0.0931 | 0.9622 | 0.3828 | 0.0139 | | 0.0012 | 0.5935 | 1.2609 |
| Alt | 25 | 0.343 | M0 | -0.0147 | 0.1104 | 0.9492 | 0.4285 | | 0.4093 | 0.0049 | 0.0000 | 1.0000 |

| Truth | nE | pE(obs) | Method | Bias | RMSE | Coverage | Mean width | Type I error | Power | MCSE | Hist. wt. | ER |
|---|---|---|---|---|---|---|---|---|---|---|---|---|
| Alt | 25 | 0.343 | M1 | -0.0290 | 0.0975 | 0.9428 | 0.3650 | | 0.4565 | 0.0050 | | 1.2836 |
| Alt | 25 | 0.343 | M2 | -0.0197 | 0.1010 | 0.9550 | 0.4020 | | 0.4208 | 0.0049 | 0.6474 | 1.1955 |
| Alt | 25 | 0.343 | M3 | -0.0214 | 0.0991 | 0.9571 | 0.3957 | | 0.4215 | 0.0049 | 0.5939 | 1.2402 |
| Null | 50 | 0.343 | M0 | -0.0087 | 0.1057 | 0.9518 | 0.4169 | 0.0220 | | 0.0015 | 0.0000 | 1.0000 |
| Null | 50 | 0.343 | M1 | -0.0254 | 0.0849 | 0.9421 | 0.3215 | 0.0115 | | 0.0011 | | 1.5497 |
| Null | 50 | 0.343 | M2 | -0.0155 | 0.0911 | 0.9598 | 0.3748 | 0.0125 | | 0.0011 | 0.6728 | 1.3449 |
| Null | 50 | 0.343 | M3 | -0.0127 | 0.0898 | 0.9607 | 0.3732 | 0.0132 | | 0.0011 | 0.6188 | 1.3848 |
| Alt | 50 | 0.343 | M0 | -0.0143 | 0.1107 | 0.9498 | 0.4285 | | 0.4068 | 0.0049 | 0.0000 | 1.0000 |
| Alt | 50 | 0.343 | M1 | -0.0318 | 0.0919 | 0.9366 | 0.3370 | | 0.5034 | 0.0050 | | 1.4516 |
| Alt | 50 | 0.343 | M2 | -0.0213 | 0.0971 | 0.9556 | 0.3891 | | 0.4468 | 0.0050 | 0.6686 | 1.2997 |
| Alt | 50 | 0.343 | M3 | -0.0187 | 0.0956 | 0.9594 | 0.3865 | | 0.4525 | 0.0050 | 0.6165 | 1.3394 |
| Null | 100 | 0.343 | M0 | -0.0080 | 0.1057 | 0.9526 | 0.4169 | 0.0231 | | 0.0015 | 0.0000 | 1.0000 |
| Null | 100 | 0.343 | M1 | -0.0296 | 0.0807 | 0.9334 | 0.2944 | 0.0099 | | 0.0010 | | 1.7137 |
| Null | 100 | 0.343 | M2 | -0.0175 | 0.0874 | 0.9597 | 0.3629 | 0.0127 | | 0.0011 | 0.6839 | 1.4605 |
| Null | 100 | 0.343 | M3 | -0.0105 | 0.0859 | 0.9635 | 0.3635 | 0.0145 | | 0.0012 | 0.6358 | 1.5126 |
| Alt | 100 | 0.343 | M0 | -0.0147 | 0.1097 | 0.9497 | 0.4286 | | 0.4041 | 0.0049 | 0.0000 | 1.0000 |
| Alt | 100 | 0.343 | M1 | -0.0365 | 0.0877 | 0.9270 | 0.3116 | | 0.5531 | 0.0050 | | 1.5664 |
| Alt | 100 | 0.343 | M2 | -0.0243 | 0.0930 | 0.9575 | 0.3771 | | 0.4665 | 0.0050 | 0.6836 | 1.3934 |
| Alt | 100 | 0.343 | M3 | -0.0174 | 0.0910 | 0.9628 | 0.3770 | | 0.4805 | 0.0050 | 0.6353 | 1.4530 |
| Null | 250 | 0.343 | M0 | -0.0073 | 0.1057 | 0.9521 | 0.4167 | 0.0213 | | 0.0014 | 0.0000 | 1.0000 |
| Null | 250 | 0.343 | M1 | -0.0327 | 0.0756 | 0.9239 | 0.2688 | 0.0070 | | 0.0008 | | 1.9536 |
| Null | 250 | 0.343 | M2 | -0.0194 | 0.0832 | 0.9604 | 0.3509 | 0.0090 | | 0.0009 | 0.6943 | 1.6125 |
| Null | 250 | 0.343 | M3 | -0.0080 | 0.0814 | 0.9684 | 0.3526 | 0.0136 | | 0.0012 | 0.6523 | 1.6866 |

| Truth | nE | pE(obs) | Method | Bias | RMSE | Coverage | Mean width | Type I error | Power | MCSE | Hist. wt. | ER |
|---|---|---|---|---|---|---|---|---|---|---|---|---|
| Alt | 250 | 0.343 | M0 | -0.0144 | 0.1100 | 0.9465 | 0.4284 | | 0.4035 | 0.0049 | 0.0000 | 1.0000 |
| Alt | 250 | 0.343 | M1 | -0.0395 | 0.0839 | 0.9195 | 0.2877 | | 0.6016 | 0.0049 | | 1.7185 |
| Alt | 250 | 0.343 | M2 | -0.0264 | 0.0906 | 0.9561 | 0.3657 | | 0.4935 | 0.0050 | 0.6944 | 1.4749 |
| Alt | 250 | 0.343 | M3 | -0.0146 | 0.0870 | 0.9637 | 0.3660 | | 0.5168 | 0.0050 | 0.6542 | 1.5977 |

*Note: M3 applies the same RMP separately within X=0 and X=1 and then standardizes the stratum-specific posterior quantities to the target trial population's 50%/50% covariate distribution.*

## Supplementary Table S4. World 3 - Qualified external data with residual outcome drift

| Truth | nE | pE(obs) | Method | Bias | RMSE | Coverage | Mean width | Type I error | Power | MCSE | Hist. wt. | ER |
|---|---|---|---|---|---|---|---|---|---|---|---|---|
| Null | 50 | 0.150 | M0 | -0.0071 | 0.1064 | 0.9486 | 0.4168 | 0.0236 | | 0.0015 | 0.0000 | 1.0000 |
| Null | 50 | 0.150 | M1 | 0.1000 | 0.1261 | 0.7449 | 0.3039 | 0.2683 | | 0.0044 | | 0.7120 |
| Null | 50 | 0.150 | M2 | 0.0307 | 0.1124 | 0.9242 | 0.4075 | 0.0662 | | 0.0025 | 0.5005 | 0.8960 |
| Null | 50 | 0.150 | M3 | 0.0307 | 0.1124 | 0.9242 | 0.4075 | 0.0662 | | 0.0025 | 0.5005 | 0.8960 |
| Alt | 50 | 0.150 | M0 | -0.0147 | 0.1111 | 0.9503 | 0.4286 | | 0.4077 | 0.0049 | 0.0000 | 1.0000 |
| Alt | 50 | 0.150 | M1 | 0.0931 | 0.1239 | 0.7969 | 0.3202 | | 0.9498 | 0.0022 | | 0.8036 |
| Alt | 50 | 0.150 | M2 | 0.0229 | 0.1144 | 0.9336 | 0.4197 | | 0.5268 | 0.0050 | 0.4982 | 0.9435 |
| Alt | 50 | 0.150 | M3 | 0.0229 | 0.1144 | 0.9336 | 0.4197 | | 0.5268 | 0.0050 | 0.4982 | 0.9435 |
| Null | 100 | 0.150 | M0 | -0.0059 | 0.1052 | 0.9573 | 0.4166 | 0.0218 | | 0.0015 | 0.0000 | 1.0000 |
| Null | 100 | 0.150 | M1 | 0.1235 | 0.1422 | 0.5767 | 0.2792 | 0.4534 | | 0.0050 | | 0.5477 |
| Null | 100 | 0.150 | M2 | 0.0363 | 0.1165 | 0.9083 | 0.4121 | 0.0853 | | 0.0028 | 0.4728 | 0.8159 |
| Null | 100 | 0.150 | M3 | 0.0363 | 0.1165 | 0.9083 | 0.4121 | 0.0853 | | 0.0028 | 0.4728 | 0.8159 |
| Alt | 100 | 0.150 | M0 | -0.0139 | 0.1095 | 0.9540 | 0.4286 | | 0.4124 | 0.0049 | 0.0000 | 1.0000 |
| Alt | 100 | 0.150 | M1 | 0.1158 | 0.1383 | 0.6754 | 0.2974 | | 0.9898 | 0.0010 | | 0.6264 |
| Alt | 100 | 0.150 | M2 | 0.0281 | 0.1177 | 0.9206 | 0.4237 | | 0.5301 | 0.0050 | 0.4724 | 0.8645 |
| Alt | 100 | 0.150 | M3 | 0.0281 | 0.1177 | 0.9206 | 0.4237 | | 0.5301 | 0.0050 | 0.4724 | 0.8645 |

| Truth | nE | pE(obs) | Method | Bias | RMSE | Coverage | Mean width | Type I error | Power | MCSE | Hist. wt. | ER |
|---|---|---|---|---|---|---|---|---|---|---|---|---|
| Null | 250 | 0.150 | M0 | -0.0062 | 0.1056 | 0.9507 | 0.4165 | 0.0234 |  | 0.0015 | 0.0000 | 1.0000 |
| Null | 250 | 0.150 | M1 | 0.1421 | 0.1569 | 0.3904 | 0.2598 | 0.6274 |  | 0.0048 |  | 0.4530 |
| Null | 250 | 0.150 | M2 | 0.0385 | 0.1213 | 0.8924 | 0.4157 | 0.1002 |  | 0.0030 | 0.4495 | 0.7584 |
| Null | 250 | 0.150 | M3 | 0.0385 | 0.1213 | 0.8924 | 0.4157 | 0.1002 |  | 0.0030 | 0.4495 | 0.7584 |
| Alt | 250 | 0.150 | M0 | -0.0164 | 0.1107 | 0.9509 | 0.4287 |  | 0.4000 | 0.0049 | 0.0000 | 1.0000 |
| Alt | 250 | 0.150 | M1 | 0.1324 | 0.1507 | 0.5472 | 0.2793 |  | 0.9982 | 0.0004 |  | 0.5394 |
| Alt | 250 | 0.150 | M2 | 0.0283 | 0.1220 | 0.9083 | 0.4285 |  | 0.5137 | 0.0050 | 0.4466 | 0.8225 |
| Alt | 250 | 0.150 | M3 | 0.0283 | 0.1220 | 0.9083 | 0.4285 |  | 0.5137 | 0.0050 | 0.4466 | 0.8225 |
| Null | 50 | 0.200 | M0 | -0.0069 | 0.1055 | 0.9517 | 0.4166 | 0.0223 |  | 0.0015 | 0.0000 | 1.0000 |
| Null | 50 | 0.200 | M1 | 0.0671 | 0.1034 | 0.8700 | 0.3095 | 0.1393 |  | 0.0035 |  | 1.0418 |
| Null | 50 | 0.200 | M2 | 0.0261 | 0.1015 | 0.9430 | 0.3901 | 0.0490 |  | 0.0022 | 0.6092 | 1.0812 |
| Null | 50 | 0.200 | M3 | 0.0261 | 0.1015 | 0.9430 | 0.3901 | 0.0490 |  | 0.0022 | 0.6092 | 1.0812 |
| Alt | 50 | 0.200 | M0 | -0.0145 | 0.1079 | 0.9557 | 0.4286 |  | 0.4027 | 0.0049 | 0.0000 | 1.0000 |
| Alt | 50 | 0.200 | M1 | 0.0592 | 0.1017 | 0.8947 | 0.3256 |  | 0.8823 | 0.0032 |  | 1.1263 |
| Alt | 50 | 0.200 | M2 | 0.0186 | 0.1014 | 0.9515 | 0.4025 |  | 0.5526 | 0.0050 | 0.6098 | 1.1324 |
| Alt | 50 | 0.200 | M3 | 0.0186 | 0.1014 | 0.9515 | 0.4025 |  | 0.5526 | 0.0050 | 0.6098 | 1.1324 |
| Null | 100 | 0.200 | M0 | -0.0071 | 0.1060 | 0.9502 | 0.4163 | 0.0233 |  | 0.0015 | 0.0000 | 1.0000 |
| Null | 100 | 0.200 | M1 | 0.0827 | 0.1097 | 0.7994 | 0.2841 | 0.2203 |  | 0.0041 |  | 0.9333 |
| Null | 100 | 0.200 | M2 | 0.0323 | 0.1040 | 0.9323 | 0.3877 | 0.0622 |  | 0.0024 | 0.6046 | 1.0370 |
| Null | 100 | 0.200 | M3 | 0.0323 | 0.1040 | 0.9323 | 0.3877 | 0.0622 |  | 0.0024 | 0.6046 | 1.0370 |
| Alt | 100 | 0.200 | M0 | -0.0135 | 0.1092 | 0.9535 | 0.4285 |  | 0.4098 | 0.0049 | 0.0000 | 1.0000 |
| Alt | 100 | 0.200 | M1 | 0.0770 | 0.1092 | 0.8392 | 0.3020 |  | 0.9543 | 0.0021 |  | 1.0002 |
| Alt | 100 | 0.200 | M2 | 0.0257 | 0.1049 | 0.9425 | 0.4010 |  | 0.5789 | 0.0049 | 0.6016 | 1.0829 |

| Truth | nE | pE(obs) | Method | Bias | RMSE | Coverage | Mean width | Type I error | Power | MCSE | Hist. wt. | ER |
|---|---|---|---|---|---|---|---|---|---|---|---|---|
| Alt | 100 | 0.200 | M3 | 0.0257 | 0.1049 | 0.9425 | 0.4010 | | 0.5789 | 0.0049 | 0.6016 | 1.0829 |
| Null | 250 | 0.200 | M0 | -0.0067 | 0.1065 | 0.9480 | 0.4169 | 0.0250 | | 0.0016 | 0.0000 | 1.0000 |
| Null | 250 | 0.200 | M1 | 0.0975 | 0.1187 | 0.6908 | 0.2629 | 0.3255 | | 0.0047 | | 0.8053 |
| Null | 250 | 0.200 | M2 | 0.0382 | 0.1077 | 0.9202 | 0.3879 | 0.0754 | | 0.0026 | 0.5968 | 0.9772 |
| Null | 250 | 0.200 | M3 | 0.0382 | 0.1077 | 0.9202 | 0.3879 | 0.0754 | | 0.0026 | 0.5968 | 0.9772 |
| Alt | 250 | 0.200 | M0 | -0.0152 | 0.1104 | 0.9484 | 0.4284 | | 0.4033 | 0.0049 | 0.0000 | 1.0000 |
| Alt | 250 | 0.200 | M1 | 0.0871 | 0.1125 | 0.7802 | 0.2822 | | 0.9868 | 0.0011 | | 0.9619 |
| Alt | 250 | 0.200 | M2 | 0.0284 | 0.1070 | 0.9353 | 0.3987 | | 0.5894 | 0.0049 | 0.5989 | 1.0630 |
| Alt | 250 | 0.200 | M3 | 0.0284 | 0.1070 | 0.9353 | 0.3987 | | 0.5894 | 0.0049 | 0.5989 | 1.0630 |
| Null | 50 | 0.250 | M0 | -0.0077 | 0.1046 | 0.9543 | 0.4170 | 0.0216 | | 0.0015 | 0.0000 | 1.0000 |
| Null | 50 | 0.250 | M1 | 0.0354 | 0.0867 | 0.9337 | 0.3144 | 0.0648 | | 0.0025 | | 1.4555 |
| Null | 50 | 0.250 | M2 | 0.0142 | 0.0921 | 0.9592 | 0.3777 | 0.0328 | | 0.0018 | 0.6688 | 1.2894 |
| Null | 50 | 0.250 | M3 | 0.0142 | 0.0921 | 0.9592 | 0.3777 | 0.0328 | | 0.0018 | 0.6688 | 1.2894 |
| Alt | 50 | 0.250 | M0 | -0.0130 | 0.1094 | 0.9507 | 0.4283 | | 0.4096 | 0.0049 | 0.0000 | 1.0000 |
| Alt | 50 | 0.250 | M1 | 0.0281 | 0.0887 | 0.9396 | 0.3302 | | 0.7804 | 0.0041 | | 1.5214 |
| Alt | 50 | 0.250 | M2 | 0.0082 | 0.0957 | 0.9581 | 0.3898 | | 0.5561 | 0.0050 | 0.6707 | 1.3073 |
| Alt | 50 | 0.250 | M3 | 0.0082 | 0.0957 | 0.9581 | 0.3898 | | 0.5561 | 0.0050 | 0.6707 | 1.3073 |
| Null | 100 | 0.250 | M0 | -0.0064 | 0.1045 | 0.9535 | 0.4166 | 0.0237 | | 0.0015 | 0.0000 | 1.0000 |
| Null | 100 | 0.250 | M1 | 0.0445 | 0.0853 | 0.9129 | 0.2885 | 0.0906 | | 0.0029 | | 1.5016 |
| Null | 100 | 0.250 | M2 | 0.0199 | 0.0901 | 0.9574 | 0.3680 | 0.0367 | | 0.0019 | 0.6799 | 1.3458 |
| Null | 100 | 0.250 | M3 | 0.0199 | 0.0901 | 0.9574 | 0.3680 | 0.0367 | | 0.0019 | 0.6799 | 1.3458 |
| Alt | 100 | 0.250 | M0 | -0.0138 | 0.1102 | 0.9502 | 0.4287 | | 0.4082 | 0.0049 | 0.0000 | 1.0000 |
| Alt | 100 | 0.250 | M1 | 0.0378 | 0.0873 | 0.9238 | 0.3059 | | 0.8654 | 0.0034 | | 1.5944 |

| Truth | nE | pE(obs) | Method | Bias | RMSE | Coverage | Mean width | Type I error | Power | MCSE | Hist. wt. | ER |
|---|---|---|---|---|---|---|---|---|---|---|---|---|
| Alt | 100 | 0.250 | M2 | 0.0129 | 0.0944 | 0.9584 | 0.3817 | | 0.5900 | 0.0049 | 0.6793 | 1.3637 |
| Alt | 100 | 0.250 | M3 | 0.0129 | 0.0944 | 0.9584 | 0.3817 | | 0.5900 | 0.0049 | 0.6793 | 1.3637 |
| Null | 250 | 0.250 | M0 | -0.0071 | 0.1057 | 0.9529 | 0.4168 | 0.0223 | | 0.0015 | 0.0000 | 1.0000 |
| Null | 250 | 0.250 | M1 | 0.0516 | 0.0850 | 0.8854 | 0.2654 | 0.1197 | | 0.0032 | | 1.5463 |
| Null | 250 | 0.250 | M2 | 0.0233 | 0.0891 | 0.9532 | 0.3607 | 0.0417 | | 0.0020 | 0.6862 | 1.4070 |
| Null | 250 | 0.250 | M3 | 0.0233 | 0.0891 | 0.9532 | 0.3607 | 0.0417 | | 0.0020 | 0.6862 | 1.4070 |
| Alt | 250 | 0.250 | M0 | -0.0144 | 0.1104 | 0.9503 | 0.4288 | | 0.4014 | 0.0049 | 0.0000 | 1.0000 |
| Alt | 250 | 0.250 | M1 | 0.0447 | 0.0858 | 0.9082 | 0.2845 | | 0.9343 | 0.0025 | | 1.6539 |
| Alt | 250 | 0.250 | M2 | 0.0164 | 0.0926 | 0.9554 | 0.3740 | | 0.6115 | 0.0049 | 0.6867 | 1.4198 |
| Alt | 250 | 0.250 | M3 | 0.0164 | 0.0926 | 0.9554 | 0.3740 | | 0.6115 | 0.0049 | 0.6867 | 1.4198 |
| Null | 50 | 0.300 | M0 | -0.0085 | 0.1046 | 0.9524 | 0.4166 | 0.0214 | | 0.0014 | 0.0000 | 1.0000 |
| Null | 50 | 0.300 | M1 | 0.0010 | 0.0816 | 0.9463 | 0.3182 | 0.0293 | | 0.0017 | | 1.6439 |
| Null | 50 | 0.300 | M2 | -0.0025 | 0.0887 | 0.9625 | 0.3722 | 0.0196 | | 0.0014 | 0.6880 | 1.3915 |
| Null | 50 | 0.300 | M3 | -0.0025 | 0.0887 | 0.9625 | 0.3722 | 0.0196 | | 0.0014 | 0.6880 | 1.3915 |
| Alt | 50 | 0.300 | M0 | -0.0154 | 0.1099 | 0.9495 | 0.4287 | | 0.4013 | 0.0049 | 0.0000 | 1.0000 |
| Alt | 50 | 0.300 | M1 | -0.0052 | 0.0854 | 0.9496 | 0.3341 | | 0.6265 | 0.0048 | | 1.6558 |
| Alt | 50 | 0.300 | M2 | -0.0090 | 0.0937 | 0.9602 | 0.3857 | | 0.5023 | 0.0050 | 0.6883 | 1.3762 |
| Alt | 50 | 0.300 | M3 | -0.0090 | 0.0937 | 0.9602 | 0.3857 | | 0.5023 | 0.0050 | 0.6883 | 1.3762 |
| Null | 100 | 0.300 | M0 | -0.0075 | 0.1059 | 0.9518 | 0.4168 | 0.0239 | | 0.0015 | 0.0000 | 1.0000 |
| Null | 100 | 0.300 | M1 | 0.0047 | 0.0735 | 0.9524 | 0.2920 | 0.0287 | | 0.0017 | | 2.0748 |
| Null | 100 | 0.300 | M2 | 0.0001 | 0.0842 | 0.9678 | 0.3601 | 0.0178 | | 0.0013 | 0.7012 | 1.5807 |
| Null | 100 | 0.300 | M3 | 0.0001 | 0.0842 | 0.9678 | 0.3601 | 0.0178 | | 0.0013 | 0.7012 | 1.5807 |
| Alt | 100 | 0.300 | M0 | -0.0166 | 0.1094 | 0.9541 | 0.4290 | | 0.3952 | 0.0049 | 0.0000 | 1.0000 |

| Truth | nE | pE(obs) | Method | Bias | RMSE | Coverage | Mean width | Type I error | Power | MCSE | Hist. wt. | ER |
|---|---|---|---|---|---|---|---|---|---|---|---|---|
| Alt | 100 | 0.300 | M1 | -0.0047 | 0.0780 | 0.9548 | 0.3094 | | 0.7078 | 0.0045 | | 1.9703 |
| Alt | 100 | 0.300 | M2 | -0.0093 | 0.0880 | 0.9663 | 0.3741 | | 0.5338 | 0.0050 | 0.7023 | 1.5466 |
| Alt | 100 | 0.300 | M3 | -0.0093 | 0.0880 | 0.9663 | 0.3741 | | 0.5338 | 0.0050 | 0.7023 | 1.5466 |
| Null | 250 | 0.300 | M0 | -0.0078 | 0.1047 | 0.9538 | 0.4167 | 0.0225 | | 0.0015 | 0.0000 | 1.0000 |
| Null | 250 | 0.300 | M1 | 0.0059 | 0.0686 | 0.9515 | 0.2672 | 0.0285 | | 0.0017 | | 2.3282 |
| Null | 250 | 0.300 | M2 | 0.0007 | 0.0801 | 0.9689 | 0.3466 | 0.0176 | | 0.0013 | 0.7168 | 1.7087 |
| Null | 250 | 0.300 | M3 | 0.0007 | 0.0801 | 0.9689 | 0.3466 | 0.0176 | | 0.0013 | 0.7168 | 1.7087 |
| Alt | 250 | 0.300 | M0 | -0.0149 | 0.1094 | 0.9509 | 0.4286 | | 0.4012 | 0.0049 | 0.0000 | 1.0000 |
| Alt | 250 | 0.300 | M1 | -0.0010 | 0.0733 | 0.9517 | 0.2864 | | 0.7860 | 0.0041 | | 2.2297 |
| Alt | 250 | 0.300 | M2 | -0.0064 | 0.0846 | 0.9701 | 0.3622 | | 0.5819 | 0.0049 | 0.7137 | 1.6723 |
| Alt | 250 | 0.300 | M3 | -0.0064 | 0.0846 | 0.9701 | 0.3622 | | 0.5819 | 0.0049 | 0.7137 | 1.6723 |
| Null | 50 | 0.350 | M0 | -0.0078 | 0.1046 | 0.9518 | 0.4168 | 0.0207 | | 0.0014 | 0.0000 | 1.0000 |
| Null | 50 | 0.350 | M1 | -0.0305 | 0.0869 | 0.9340 | 0.3219 | 0.0105 | | 0.0010 | | 1.4493 |
| Null | 50 | 0.350 | M2 | -0.0174 | 0.0911 | 0.9587 | 0.3759 | 0.0116 | | 0.0011 | 0.6675 | 1.3187 |
| Null | 50 | 0.350 | M3 | -0.0174 | 0.0911 | 0.9587 | 0.3759 | 0.0116 | | 0.0011 | 0.6675 | 1.3187 |
| Alt | 50 | 0.350 | M0 | -0.0130 | 0.1085 | 0.9547 | 0.4284 | | 0.4058 | 0.0049 | 0.0000 | 1.0000 |
| Alt | 50 | 0.350 | M1 | -0.0364 | 0.0930 | 0.9340 | 0.3375 | | 0.4738 | 0.0050 | | 1.3604 |
| Alt | 50 | 0.350 | M2 | -0.0224 | 0.0961 | 0.9586 | 0.3895 | | 0.4370 | 0.0050 | 0.6653 | 1.2742 |
| Alt | 50 | 0.350 | M3 | -0.0224 | 0.0961 | 0.9586 | 0.3895 | | 0.4370 | 0.0050 | 0.6653 | 1.2742 |
| Null | 100 | 0.350 | M0 | -0.0090 | 0.1068 | 0.9498 | 0.4167 | 0.0215 | | 0.0015 | 0.0000 | 1.0000 |
| Null | 100 | 0.350 | M1 | -0.0357 | 0.0834 | 0.9231 | 0.2945 | 0.0071 | | 0.0008 | | 1.6396 |
| Null | 100 | 0.350 | M2 | -0.0214 | 0.0894 | 0.9557 | 0.3634 | 0.0098 | | 0.0010 | 0.6805 | 1.4296 |
| Null | 100 | 0.350 | M3 | -0.0214 | 0.0894 | 0.9557 | 0.3634 | 0.0098 | | 0.0010 | 0.6805 | 1.4296 |

| Truth | nE | pE(obs) | Method | Bias | RMSE | Coverage | Mean width | Type I error | Power | MCSE | Hist. wt. | ER |
|---|---|---|---|---|---|---|---|---|---|---|---|---|
| Alt | 100 | 0.350 | M0 | -0.0170 | 0.1119 | 0.9446 | 0.4289 | | 0.3934 | 0.0049 | 0.0000 | 1.0000 |
| Alt | 100 | 0.350 | M1 | -0.0451 | 0.0923 | 0.9127 | 0.3120 | | 0.5067 | 0.0050 | | 1.4691 |
| Alt | 100 | 0.350 | M2 | -0.0297 | 0.0965 | 0.9498 | 0.3781 | | 0.4409 | 0.0050 | 0.6792 | 1.3451 |
| Alt | 100 | 0.350 | M3 | -0.0297 | 0.0965 | 0.9498 | 0.3781 | | 0.4409 | 0.0050 | 0.6792 | 1.3451 |
| Null | 250 | 0.350 | M0 | -0.0078 | 0.1048 | 0.9543 | 0.4164 | 0.0206 | | 0.0014 | 0.0000 | 1.0000 |
| Null | 250 | 0.350 | M1 | -0.0402 | 0.0791 | 0.9110 | 0.2688 | 0.0053 | | 0.0007 | | 1.7562 |
| Null | 250 | 0.350 | M2 | -0.0230 | 0.0850 | 0.9572 | 0.3509 | 0.0087 | | 0.0009 | 0.6896 | 1.5210 |
| Null | 250 | 0.350 | M3 | -0.0230 | 0.0850 | 0.9572 | 0.3509 | 0.0087 | | 0.0009 | 0.6896 | 1.5210 |
| Alt | 250 | 0.350 | M0 | -0.0152 | 0.1106 | 0.9488 | 0.4284 | | 0.4022 | 0.0049 | 0.0000 | 1.0000 |
| Alt | 250 | 0.350 | M1 | -0.0475 | 0.0879 | 0.9001 | 0.2881 | | 0.5636 | 0.0050 | | 1.5819 |
| Alt | 250 | 0.350 | M2 | -0.0303 | 0.0925 | 0.9504 | 0.3669 | | 0.4783 | 0.0050 | 0.6880 | 1.4299 |
| Alt | 250 | 0.350 | M3 | -0.0303 | 0.0925 | 0.9504 | 0.3669 | | 0.4783 | 0.0050 | 0.6880 | 1.4299 |
| Null | 50 | 0.400 | M0 | -0.0064 | 0.1070 | 0.9489 | 0.4164 | 0.0243 | | 0.0015 | 0.0000 | 1.0000 |
| Null | 50 | 0.400 | M1 | -0.0625 | 0.1030 | 0.8825 | 0.3248 | 0.0032 | | 0.0006 | | 1.0783 |
| Null | 50 | 0.400 | M2 | -0.0290 | 0.1000 | 0.9467 | 0.3883 | 0.0117 | | 0.0011 | 0.6045 | 1.1446 |
| Null | 50 | 0.400 | M3 | -0.0290 | 0.1000 | 0.9467 | 0.3883 | 0.0117 | | 0.0011 | 0.6045 | 1.1446 |
| Alt | 50 | 0.400 | M0 | -0.0148 | 0.1097 | 0.9504 | 0.4286 | | 0.4022 | 0.0049 | 0.0000 | 1.0000 |
| Alt | 50 | 0.400 | M1 | -0.0712 | 0.1124 | 0.8737 | 0.3404 | | 0.3220 | 0.0047 | | 0.9525 |
| Alt | 50 | 0.400 | M2 | -0.0381 | 0.1058 | 0.9430 | 0.4010 | | 0.3547 | 0.0048 | 0.6097 | 1.0750 |
| Alt | 50 | 0.400 | M3 | -0.0381 | 0.1058 | 0.9430 | 0.4010 | | 0.3547 | 0.0048 | 0.6097 | 1.0750 |
| Null | 100 | 0.400 | M0 | -0.0081 | 0.1051 | 0.9533 | 0.4167 | 0.0215 | | 0.0015 | 0.0000 | 1.0000 |
| Null | 100 | 0.400 | M1 | -0.0755 | 0.1069 | 0.8313 | 0.2967 | 0.0013 | | 0.0004 | | 0.9681 |
| Null | 100 | 0.400 | M2 | -0.0371 | 0.0996 | 0.9415 | 0.3793 | 0.0080 | | 0.0009 | 0.6147 | 1.1150 |

| Truth | nE | pE(obs) | Method | Bias | RMSE | Coverage | Mean width | Type I error | Power | MCSE | Hist. wt. | ER |
|---|---|---|---|---|---|---|---|---|---|---|---|---|
| Null | 100 | 0.400 | M3 | -0.0371 | 0.0996 | 0.9415 | 0.3793 | 0.0080 | | 0.0009 | 0.6147 | 1.1150 |
| Alt | 100 | 0.400 | M0 | -0.0138 | 0.1093 | 0.9521 | 0.4284 | | 0.4032 | 0.0049 | 0.0000 | 1.0000 |
| Alt | 100 | 0.400 | M1 | -0.0820 | 0.1148 | 0.8275 | 0.3141 | | 0.3202 | 0.0047 | | 0.9057 |
| Alt | 100 | 0.400 | M2 | -0.0424 | 0.1055 | 0.9368 | 0.3937 | | 0.3523 | 0.0048 | 0.6122 | 1.0739 |
| Alt | 100 | 0.400 | M3 | -0.0424 | 0.1055 | 0.9368 | 0.3937 | | 0.3523 | 0.0048 | 0.6122 | 1.0739 |
| Null | 250 | 0.400 | M0 | -0.0053 | 0.1055 | 0.9506 | 0.4166 | 0.0251 | | 0.0016 | 0.0000 | 1.0000 |
| Null | 250 | 0.400 | M1 | -0.0832 | 0.1080 | 0.7887 | 0.2704 | 0.0011 | | 0.0003 | | 0.9541 |
| Null | 250 | 0.400 | M2 | -0.0396 | 0.0998 | 0.9353 | 0.3716 | 0.0095 | | 0.0010 | 0.6169 | 1.1170 |
| Null | 250 | 0.400 | M3 | -0.0396 | 0.0998 | 0.9353 | 0.3716 | 0.0095 | | 0.0010 | 0.6169 | 1.1170 |
| Alt | 250 | 0.400 | M0 | -0.0132 | 0.1081 | 0.9538 | 0.4285 | | 0.4063 | 0.0049 | 0.0000 | 1.0000 |
| Alt | 250 | 0.400 | M1 | -0.0901 | 0.1159 | 0.7788 | 0.2892 | | 0.3264 | 0.0047 | | 0.8707 |
| Alt | 250 | 0.400 | M2 | -0.0465 | 0.1040 | 0.9326 | 0.3872 | | 0.3503 | 0.0048 | 0.6159 | 1.0810 |
| Alt | 250 | 0.400 | M3 | -0.0465 | 0.1040 | 0.9326 | 0.3872 | | 0.3503 | 0.0048 | 0.6159 | 1.0810 |
| Null | 50 | 0.450 | M0 | -0.0066 | 0.1063 | 0.9490 | 0.4166 | 0.0243 | | 0.0015 | 0.0000 | 1.0000 |
| Null | 50 | 0.450 | M1 | -0.0939 | 0.1246 | 0.8038 | 0.3269 | 0.0012 | | 0.0003 | | 0.7287 |
| Null | 50 | 0.450 | M2 | -0.0375 | 0.1084 | 0.9349 | 0.4021 | 0.0132 | | 0.0011 | 0.5301 | 0.9625 |
| Null | 50 | 0.450 | M3 | -0.0375 | 0.1084 | 0.9349 | 0.4021 | 0.0132 | | 0.0011 | 0.5301 | 0.9625 |
| Alt | 50 | 0.450 | M0 | -0.0149 | 0.1102 | 0.9511 | 0.4286 | | 0.4029 | 0.0049 | 0.0000 | 1.0000 |
| Alt | 50 | 0.450 | M1 | -0.1023 | 0.1353 | 0.7891 | 0.3424 | | 0.2056 | 0.0040 | | 0.6629 |
| Alt | 50 | 0.450 | M2 | -0.0463 | 0.1151 | 0.9277 | 0.4159 | | 0.3010 | 0.0046 | 0.5305 | 0.9159 |
| Alt | 50 | 0.450 | M3 | -0.0463 | 0.1151 | 0.9277 | 0.4159 | | 0.3010 | 0.0046 | 0.5305 | 0.9159 |
| Null | 100 | 0.450 | M0 | -0.0068 | 0.1049 | 0.9524 | 0.4166 | 0.0207 | | 0.0014 | 0.0000 | 1.0000 |
| Null | 100 | 0.450 | M1 | -0.1138 | 0.1366 | 0.6934 | 0.2985 | 0.0000 | | 0.0000 | | 0.5900 |

| Truth | nE | pE(obs) | Method | Bias | RMSE | Coverage | Mean width | Type I error | Power | MCSE | Hist. wt. | ER |
|---|---|---|---|---|---|---|---|---|---|---|---|---|
| Null | 100 | 0.450 | M2 | -0.0449 | 0.1112 | 0.9279 | 0.4009 | 0.0092 | | 0.0010 | 0.5185 | 0.8906 |
| Null | 100 | 0.450 | M3 | -0.0449 | 0.1112 | 0.9279 | 0.4009 | 0.0092 | | 0.0010 | 0.5185 | 0.8906 |
| Alt | 100 | 0.450 | M0 | -0.0120 | 0.1089 | 0.9510 | 0.4283 | | 0.4122 | 0.0049 | 0.0000 | 1.0000 |
| Alt | 100 | 0.450 | M1 | -0.1191 | 0.1440 | 0.6935 | 0.3154 | | 0.1737 | 0.0038 | | 0.5723 |
| Alt | 100 | 0.450 | M2 | -0.0503 | 0.1169 | 0.9227 | 0.4147 | | 0.2802 | 0.0045 | 0.5190 | 0.8681 |
| Alt | 100 | 0.450 | M3 | -0.0503 | 0.1169 | 0.9227 | 0.4147 | | 0.2802 | 0.0045 | 0.5190 | 0.8681 |
| Null | 250 | 0.450 | M0 | -0.0071 | 0.1062 | 0.9506 | 0.4167 | 0.0234 | | 0.0015 | 0.0000 | 1.0000 |
| Null | 250 | 0.450 | M1 | -0.1284 | 0.1459 | 0.5672 | 0.2711 | 0.0000 | | 0.0000 | | 0.5302 |
| Null | 250 | 0.450 | M2 | -0.0502 | 0.1154 | 0.9110 | 0.3984 | 0.0110 | | 0.0010 | 0.5125 | 0.8475 |
| Null | 250 | 0.450 | M3 | -0.0502 | 0.1154 | 0.9110 | 0.3984 | 0.0110 | | 0.0010 | 0.5125 | 0.8475 |
| Alt | 250 | 0.450 | M0 | -0.0144 | 0.1105 | 0.9520 | 0.4285 | | 0.4051 | 0.0049 | 0.0000 | 1.0000 |
| Alt | 250 | 0.450 | M1 | -0.1364 | 0.1557 | 0.5543 | 0.2898 | | 0.1436 | 0.0035 | | 0.5037 |
| Alt | 250 | 0.450 | M2 | -0.0574 | 0.1222 | 0.9094 | 0.4139 | | 0.2541 | 0.0044 | 0.5097 | 0.8176 |
| Alt | 250 | 0.450 | M3 | -0.0574 | 0.1222 | 0.9094 | 0.4139 | | 0.2541 | 0.0044 | 0.5097 | 0.8176 |

*Note: M2 and M3 are identical by design because Steps 1–4 identify no material difference requiring repair or exclusion; the external information proceeds unchanged to Step 5.*

## Supplementary Table S5. World 4 - Known endpoint mismatch

| Truth | nE | q | Method | Bias | RMSE | Coverage | Mean width | Type I error | Power | MCSE | Hist. wt. | ER |
|---|---|---|---|---|---|---|---|---|---|---|---|---|
| Null | 50 | 0.05 | M0 | -0.0078 | 0.1061 | 0.9517 | 0.4166 | 0.0210 | | 0.0014 | 0.0000 | 1.0000 |
| Null | 50 | 0.05 | M1 | -0.0205 | 0.0848 | 0.9380 | 0.3208 | 0.0143 | | 0.0012 | | 1.5655 |
| Null | 50 | 0.05 | M2 | -0.0127 | 0.0912 | 0.9577 | 0.3737 | 0.0135 | | 0.0012 | 0.6772 | 1.3537 |
| Null | 50 | 0.05 | M3 | -0.0078 | 0.1061 | 0.9517 | 0.4166 | 0.0210 | | 0.0014 | 0.0000 | 1.0000 |
| Alt | 50 | 0.05 | M0 | -0.0128 | 0.1104 | 0.9479 | 0.4286 | | 0.4141 | 0.0049 | 0.0000 | 1.0000 |
| Alt | 50 | 0.05 | M1 | -0.0256 | 0.0906 | 0.9388 | 0.3365 | | 0.5333 | 0.0050 | | 1.4856 |

| Truth | nE | q | Method | Bias | RMSE | Coverage | Mean width | Type I error | Power | MCSE | Hist. wt. | ER |
|---|---|---|---|---|---|---|---|---|---|---|---|---|
| Alt | 50 | 0.05 | M2 | -0.0177 | 0.0961 | 0.9546 | 0.3876 | | 0.4696 | 0.0050 | 0.6770 | 1.3205 |
| Alt | 50 | 0.05 | M3 | -0.0128 | 0.1104 | 0.9479 | 0.4286 | | 0.4141 | 0.0049 | 0.0000 | 1.0000 |
| Null | 100 | 0.05 | M0 | -0.0060 | 0.1066 | 0.9510 | 0.4166 | 0.0232 | | 0.0015 | 0.0000 | 1.0000 |
| Null | 100 | 0.05 | M1 | -0.0227 | 0.0788 | 0.9370 | 0.2941 | 0.0121 | | 0.0011 | | 1.8312 |
| Null | 100 | 0.05 | M2 | -0.0131 | 0.0875 | 0.9608 | 0.3610 | 0.0129 | | 0.0011 | 0.6901 | 1.4836 |
| Null | 100 | 0.05 | M3 | -0.0060 | 0.1066 | 0.9510 | 0.4166 | 0.0232 | | 0.0015 | 0.0000 | 1.0000 |
| Alt | 100 | 0.05 | M0 | -0.0150 | 0.1102 | 0.9521 | 0.4288 | | 0.4057 | 0.0049 | 0.0000 | 1.0000 |
| Alt | 100 | 0.05 | M1 | -0.0306 | 0.0854 | 0.9341 | 0.3112 | | 0.5791 | 0.0049 | | 1.6661 |
| Alt | 100 | 0.05 | M2 | -0.0217 | 0.0924 | 0.9572 | 0.3753 | | 0.4806 | 0.0050 | 0.6905 | 1.4225 |
| Alt | 100 | 0.05 | M3 | -0.0150 | 0.1102 | 0.9521 | 0.4288 | | 0.4057 | 0.0049 | 0.0000 | 1.0000 |
| Null | 250 | 0.05 | M0 | -0.0065 | 0.1061 | 0.9501 | 0.4165 | 0.0229 | | 0.0015 | 0.0000 | 1.0000 |
| Null | 250 | 0.05 | M1 | -0.0253 | 0.0731 | 0.9336 | 0.2686 | 0.0097 | | 0.0010 | | 2.1067 |
| Null | 250 | 0.05 | M2 | -0.0149 | 0.0827 | 0.9602 | 0.3487 | 0.0115 | | 0.0011 | 0.7012 | 1.6441 |
| Null | 250 | 0.05 | M3 | -0.0065 | 0.1061 | 0.9501 | 0.4165 | 0.0229 | | 0.0015 | 0.0000 | 1.0000 |
| Alt | 250 | 0.05 | M0 | -0.0146 | 0.1103 | 0.9502 | 0.4285 | | 0.4023 | 0.0049 | 0.0000 | 1.0000 |
| Alt | 250 | 0.05 | M1 | -0.0329 | 0.0809 | 0.9307 | 0.2875 | | 0.6281 | 0.0048 | | 1.8583 |
| Alt | 250 | 0.05 | M2 | -0.0229 | 0.0895 | 0.9598 | 0.3633 | | 0.5108 | 0.0050 | 0.7015 | 1.5185 |
| Alt | 250 | 0.05 | M3 | -0.0146 | 0.1103 | 0.9502 | 0.4285 | | 0.4023 | 0.0049 | 0.0000 | 1.0000 |
| Null | 50 | 0.10 | M0 | -0.0069 | 0.1051 | 0.9555 | 0.4161 | 0.0216 | | 0.0015 | 0.0000 | 1.0000 |
| Null | 50 | 0.10 | M1 | -0.0433 | 0.0922 | 0.9211 | 0.3230 | 0.0064 | | 0.0008 | | 1.2989 |
| Null | 50 | 0.10 | M2 | -0.0222 | 0.0937 | 0.9584 | 0.3802 | 0.0104 | | 0.0010 | 0.6438 | 1.2583 |
| Null | 50 | 0.10 | M3 | -0.0069 | 0.1051 | 0.9555 | 0.4161 | 0.0216 | | 0.0015 | 0.0000 | 1.0000 |
| Alt | 50 | 0.10 | M0 | -0.0161 | 0.1090 | 0.9509 | 0.4289 | | 0.3918 | 0.0049 | 0.0000 | 1.0000 |

| Truth | nE | q | Method | Bias | RMSE | Coverage | Mean width | Type I error | Power | MCSE | Hist. wt. | ER |
|---|---|---|---|---|---|---|---|---|---|---|---|---|
| Alt | 50 | 0.10 | M1 | -0.0517 | 0.1005 | 0.9118 | 0.3389 |  | 0.4088 | 0.0049 |  | 1.1775 |
| Alt | 50 | 0.10 | M2 | -0.0314 | 0.1002 | 0.9522 | 0.3933 |  | 0.3945 | 0.0049 | 0.6487 | 1.1834 |
| Alt | 50 | 0.10 | M3 | -0.0161 | 0.1090 | 0.9509 | 0.4289 |  | 0.3918 | 0.0049 | 0.0000 | 1.0000 |
| Null | 100 | 0.10 | M0 | -0.0062 | 0.1059 | 0.9505 | 0.4163 | 0.0236 |  | 0.0015 | 0.0000 | 1.0000 |
| Null | 100 | 0.10 | M1 | -0.0500 | 0.0907 | 0.9014 | 0.2955 | 0.0052 |  | 0.0007 |  | 1.3643 |
| Null | 100 | 0.10 | M2 | -0.0261 | 0.0923 | 0.9514 | 0.3687 | 0.0102 |  | 0.0010 | 0.6574 | 1.3182 |
| Null | 100 | 0.10 | M3 | -0.0062 | 0.1059 | 0.9505 | 0.4163 | 0.0236 |  | 0.0015 | 0.0000 | 1.0000 |
| Alt | 100 | 0.10 | M0 | -0.0146 | 0.1100 | 0.9509 | 0.4287 |  | 0.4044 | 0.0049 | 0.0000 | 1.0000 |
| Alt | 100 | 0.10 | M1 | -0.0577 | 0.0983 | 0.8898 | 0.3129 |  | 0.4399 | 0.0050 |  | 1.2537 |
| Alt | 100 | 0.10 | M2 | -0.0345 | 0.0982 | 0.9494 | 0.3830 |  | 0.4088 | 0.0049 | 0.6604 | 1.2550 |
| Alt | 100 | 0.10 | M3 | -0.0146 | 0.1100 | 0.9509 | 0.4287 |  | 0.4044 | 0.0049 | 0.0000 | 1.0000 |
| Null | 250 | 0.10 | M0 | -0.0088 | 0.1068 | 0.9487 | 0.4167 | 0.0231 |  | 0.0015 | 0.0000 | 1.0000 |
| Null | 250 | 0.10 | M1 | -0.0583 | 0.0906 | 0.8665 | 0.2693 | 0.0020 |  | 0.0004 |  | 1.3878 |
| Null | 250 | 0.10 | M2 | -0.0317 | 0.0920 | 0.9458 | 0.3579 | 0.0077 |  | 0.0009 | 0.6650 | 1.3462 |
| Null | 250 | 0.10 | M3 | -0.0088 | 0.1068 | 0.9487 | 0.4167 | 0.0231 |  | 0.0015 | 0.0000 | 1.0000 |
| Alt | 250 | 0.10 | M0 | -0.0150 | 0.1096 | 0.9533 | 0.4287 |  | 0.4023 | 0.0049 | 0.0000 | 1.0000 |
| Alt | 250 | 0.10 | M1 | -0.0643 | 0.0982 | 0.8636 | 0.2884 |  | 0.4639 | 0.0050 |  | 1.2478 |
| Alt | 250 | 0.10 | M2 | -0.0381 | 0.0969 | 0.9450 | 0.3735 |  | 0.4223 | 0.0049 | 0.6660 | 1.2794 |
| Alt | 250 | 0.10 | M3 | -0.0150 | 0.1096 | 0.9533 | 0.4287 |  | 0.4023 | 0.0049 | 0.0000 | 1.0000 |
| Null | 50 | 0.20 | M0 | -0.0066 | 0.1054 | 0.9523 | 0.4166 | 0.0218 |  | 0.0015 | 0.0000 | 1.0000 |
| Null | 50 | 0.20 | M1 | -0.0883 | 0.1210 | 0.8188 | 0.3265 | 0.0014 |  | 0.0004 |  | 0.7589 |
| Null | 50 | 0.20 | M2 | -0.0368 | 0.1063 | 0.9394 | 0.3999 | 0.0113 |  | 0.0011 | 0.5457 | 0.9836 |
| Null | 50 | 0.20 | M3 | -0.0066 | 0.1054 | 0.9523 | 0.4166 | 0.0218 |  | 0.0015 | 0.0000 | 1.0000 |

| Truth | nE | q | Method | Bias | RMSE | Coverage | Mean width | Type I error | Power | MCSE | Hist. wt. | ER |
|---|---|---|---|---|---|---|---|---|---|---|---|---|
| Alt | 50 | 0.20 | M0 | -0.0137 | 0.1090 | 0.9521 | 0.4284 | | 0.4057 | 0.0049 | 0.0000 | 1.0000 |
| Alt | 50 | 0.20 | M1 | -0.0958 | 0.1289 | 0.8081 | 0.3420 | | 0.2255 | 0.0042 | | 0.7160 |
| Alt | 50 | 0.20 | M2 | -0.0436 | 0.1119 | 0.9349 | 0.4134 | | 0.3154 | 0.0046 | 0.5430 | 0.9491 |
| Alt | 50 | 0.20 | M3 | -0.0137 | 0.1090 | 0.9521 | 0.4284 | | 0.4057 | 0.0049 | 0.0000 | 1.0000 |
| Null | 100 | 0.20 | M0 | -0.0060 | 0.1052 | 0.9526 | 0.4167 | 0.0238 | | 0.0015 | 0.0000 | 1.0000 |
| Null | 100 | 0.20 | M1 | -0.1049 | 0.1294 | 0.7272 | 0.2982 | 0.0003 | | 0.0002 | | 0.6613 |
| Null | 100 | 0.20 | M2 | -0.0431 | 0.1092 | 0.9279 | 0.3962 | 0.0116 | | 0.0011 | 0.5411 | 0.9287 |
| Null | 100 | 0.20 | M3 | -0.0060 | 0.1052 | 0.9526 | 0.4167 | 0.0238 | | 0.0015 | 0.0000 | 1.0000 |
| Alt | 100 | 0.20 | M0 | -0.0149 | 0.1095 | 0.9503 | 0.4285 | | 0.4006 | 0.0049 | 0.0000 | 1.0000 |
| Alt | 100 | 0.20 | M1 | -0.1140 | 0.1397 | 0.7051 | 0.3152 | | 0.1941 | 0.0040 | | 0.6138 |
| Alt | 100 | 0.20 | M2 | -0.0519 | 0.1158 | 0.9242 | 0.4109 | | 0.2781 | 0.0045 | 0.5392 | 0.8943 |
| Alt | 100 | 0.20 | M3 | -0.0149 | 0.1095 | 0.9503 | 0.4285 | | 0.4006 | 0.0049 | 0.0000 | 1.0000 |
| Null | 250 | 0.20 | M0 | -0.0063 | 0.1064 | 0.9526 | 0.4161 | 0.0227 | | 0.0015 | 0.0000 | 1.0000 |
| Null | 250 | 0.20 | M1 | -0.1208 | 0.1391 | 0.6081 | 0.2707 | 0.0002 | | 0.0001 | | 0.5856 |
| Null | 250 | 0.20 | M2 | -0.0479 | 0.1128 | 0.9183 | 0.3941 | 0.0100 | | 0.0010 | 0.5273 | 0.8908 |
| Null | 250 | 0.20 | M3 | -0.0063 | 0.1064 | 0.9526 | 0.4161 | 0.0227 | | 0.0015 | 0.0000 | 1.0000 |
| Alt | 250 | 0.20 | M0 | -0.0133 | 0.1099 | 0.9505 | 0.4283 | | 0.4113 | 0.0049 | 0.0000 | 1.0000 |
| Alt | 250 | 0.20 | M1 | -0.1272 | 0.1469 | 0.6041 | 0.2898 | | 0.1696 | 0.0038 | | 0.5603 |
| Alt | 250 | 0.20 | M2 | -0.0552 | 0.1184 | 0.9086 | 0.4086 | | 0.2694 | 0.0044 | 0.5304 | 0.8617 |
| Alt | 250 | 0.20 | M3 | -0.0133 | 0.1099 | 0.9505 | 0.4283 | | 0.4113 | 0.0049 | 0.0000 | 1.0000 |

*Note: M3 is identical to M0 by design because the source is not qualified for primary borrowing and no external information proceeds to Step 5.*

## Supplementary Table S6. World 5 - Same outcome distribution, different QTB decisions

| Setting | QTB status | nE | pE(obs) | Method | Bias | RMSE | Coverage | Mean width | Type I error | MCSE | Hist. wt. | ER |
|---|---|---|---|---|---|---|---|---|---|---|---|---|

| Setting | QTB status | nE | pE(obs) | Method | Bias | RMSE | Coverage | Mean width | Type I error | MCSE | Hist. wt. | ER |
|---|---|---|---|---|---|---|---|---|---|---|---|---|
| Residual drift | Qualified | 50 | 0.37 | M0 | -0.0089 | 0.1064 | 0.9484 | 0.4169 | 0.0226 | 0.0015 | 0.0000 | 1.0000 |
| Residual drift | Qualified | 50 | 0.37 | M1 | -0.0447 | 0.0942 | 0.9132 | 0.3231 | 0.0076 | 0.0009 | | 1.2757 |
| Residual drift | Qualified | 50 | 0.37 | M2 | -0.0241 | 0.0953 | 0.9505 | 0.3807 | 0.0112 | 0.0011 | 0.6447 | 1.2464 |
| Residual drift | Qualified | 50 | 0.37 | M3 | -0.0241 | 0.0953 | 0.9505 | 0.3807 | 0.0112 | 0.0011 | 0.6447 | 1.2464 |
| Endpoint mismatch | Not qualified | 50 | 0.37 | M0 | -0.0063 | 0.1053 | 0.9539 | 0.4162 | 0.0237 | 0.0015 | 0.0000 | 1.0000 |
| Endpoint mismatch | Not qualified | 50 | 0.37 | M1 | -0.0438 | 0.0927 | 0.9205 | 0.3231 | 0.0079 | 0.0009 | | 1.2901 |
| Endpoint mismatch | Not qualified | 50 | 0.37 | M2 | -0.0222 | 0.0943 | 0.9577 | 0.3802 | 0.0114 | 0.0011 | 0.6435 | 1.2455 |
| Endpoint mismatch | Not qualified | 50 | 0.37 | M3 | -0.0063 | 0.1053 | 0.9539 | 0.4162 | 0.0237 | 0.0015 | 0.0000 | 1.0000 |
| Residual drift | Qualified | 100 | 0.37 | M0 | -0.0073 | 0.1055 | 0.9537 | 0.4167 | 0.0206 | 0.0014 | 0.0000 | 1.0000 |
| Residual drift | Qualified | 100 | 0.37 | M1 | -0.0506 | 0.0913 | 0.8938 | 0.2957 | 0.0046 | 0.0007 | | 1.3327 |
| Residual drift | Qualified | 100 | 0.37 | M2 | -0.0269 | 0.0922 | 0.9529 | 0.3684 | 0.0082 | 0.0009 | 0.6589 | 1.3075 |
| Residual drift | Qualified | 100 | 0.37 | M3 | -0.0269 | 0.0922 | 0.9529 | 0.3684 | 0.0082 | 0.0009 | 0.6589 | 1.3075 |
| Endpoint mismatch | Not qualified | 100 | 0.37 | M0 | -0.0061 | 0.1049 | 0.9506 | 0.4166 | 0.0235 | 0.0015 | 0.0000 | 1.0000 |
| Endpoint mismatch | Not qualified | 100 | 0.37 | M1 | -0.0497 | 0.0897 | 0.9007 | 0.2957 | 0.0046 | 0.0007 | | 1.3669 |
| Endpoint mismatch | Not qualified | 100 | 0.37 | M2 | -0.0256 | 0.0912 | 0.9533 | 0.3687 | 0.0095 | 0.0010 | 0.6574 | 1.3234 |
| Endpoint mismatch | Not qualified | 100 | 0.37 | M3 | -0.0061 | 0.1049 | 0.9506 | 0.4166 | 0.0235 | 0.0015 | 0.0000 | 1.0000 |
| Residual drift | Qualified | 250 | 0.37 | M0 | -0.0053 | 0.1049 | 0.9541 | 0.4165 | 0.0222 | 0.0015 | 0.0000 | 1.0000 |
| Residual drift | Qualified | 250 | 0.37 | M1 | -0.0560 | 0.0877 | 0.8731 | 0.2697 | 0.0021 | 0.0005 | | 1.4323 |

| Setting | QTB status | nE | pE(obs) | Method | Bias | RMSE | Coverage | Mean width | Type I error | MCSE | Hist. wt. | ER |
|---|---|---|---|---|---|---|---|---|---|---|---|---|
| Residual drift | Qualified | 250 | 0.37 | M2 | -0.0286 | 0.0893 | 0.9540 | 0.3585 | 0.0067 | 0.0008 | 0.6629 | 1.3791 |
| Residual drift | Qualified | 250 | 0.37 | M3 | -0.0286 | 0.0893 | 0.9540 | 0.3585 | 0.0067 | 0.0008 | 0.6629 | 1.3791 |
| Endpoint mismatch | Not qualified | 250 | 0.37 | M0 | -0.0055 | 0.1047 | 0.9525 | 0.4165 | 0.0237 | 0.0015 | 0.0000 | 1.0000 |
| Endpoint mismatch | Not qualified | 250 | 0.37 | M1 | -0.0564 | 0.0888 | 0.8710 | 0.2696 | 0.0024 | 0.0005 | | 1.3904 |
| Endpoint mismatch | Not qualified | 250 | 0.37 | M2 | -0.0288 | 0.0899 | 0.9471 | 0.3577 | 0.0082 | 0.0009 | 0.6645 | 1.3565 |
| Endpoint mismatch | Not qualified | 250 | 0.37 | M3 | -0.0055 | 0.1047 | 0.9525 | 0.4165 | 0.0237 | 0.0015 | 0.0000 | 1.0000 |

*Note: The residual-drift and endpoint-mismatch settings have the same recorded external response probability (0.37) but different pre-borrowing actions. M3=M2 in the qualified residual-drift setting, whereas M3=M0 in the endpoint-mismatch setting that is not qualified for primary borrowing.*